\documentclass[a4paper,11pt]{article}
\pdfoutput=1
\usepackage{jheppub}
\usepackage{amsmath,amssymb,latexsym}
\usepackage{braket}
\usepackage{bm}
\usepackage{hepunits}
\usepackage[svgnames]{xcolor}

\newcommand{\nn}{\nonumber}
\newcommand{\mK}{\mathcal{K}}

\allowdisplaybreaks

\title{Two-loop infrared singularities in the production of a Higgs boson associated with a top-quark pair}

\author[b]{Jiaqi Chen,}
\author[a]{Chichuan Ma,}
\author[c,d,e]{Guoxing Wang,}
\author[c]{Li Lin Yang,}
\author[a]{Xiaoping Ye}

\affiliation[a]{School of Physics and State Key Laboratory of Nuclear Physics and Technology, Peking University, \\
No. 5 Yiheyuan Road, Beijing 100871, China}
\affiliation[b]{Institute of High Energy Physics, Chinese Academy of Sciences, \\
No. 19 Yuquan Road, Beijing 100049, China}
\affiliation[c]{Zhejiang Institute of Modern Physics, School of Physics, Zhejiang University, \\
No. 866 Yuhangtang Road, Hangzhou 310058, China}
\affiliation[d]{Institute for Theoretical Physics Amsterdam and Delta Institute for Theoretical Physics, University of Amsterdam, \\
Science Park 904, 1098 XH Amsterdam, Netherlands}
\affiliation[e]{Nikhef, Theory Group, \\
Science Park 105, 1098 XG, Amsterdam, Netherlands}

\emailAdd{chenjq@ihep.ac.cn}
\emailAdd{chichuanma@pku.edu.cn}
\emailAdd{wangguoxing2015@pku.edu.cn}
\emailAdd{yanglilin@zju.edu.cn}
\emailAdd{yexiaoping@pku.edu.cn}

\abstract{The associated production of a Higgs boson and a top-quark pair is important for probing the Yukawa coupling of the top quark, and calls for better theoretical modeling. In this paper, we calculate the two-loop infrared divergences in $t\bar{t}H$ production at hadron colliders. To do that we compute the one-loop amplitudes to higher orders in the dimensional regulator $\epsilon$. Numeric results for the infrared poles are given as a reference at several representative phase-space points. The result in this work serves as a part of the ongoing efforts towards the $t\bar{t}H$ cross sections at the next-to-next-to-leading order.}

\begin{document}

\maketitle

\clearpage

\section{Introduction}

The associated production of a top quark pair and a Higgs boson is one of the most important processes to study the Yukawa coupling of the top quark at the Large Hadron Collider (LHC) and the next-generation experimental facilities. The Yukawa coupling is crucial to understanding the origin of the large mass of the top quark. It can also probe the possible violation of the CP symmetry in the top quark sector \cite{CMS:2020cga, ATLAS:2020ior}. Such a violation is required to generate the matter-anti-matter asymmetry in our observable universe. With an integrated luminosity up to \unit{139}{\invfb}, the LHC Run 2 has measured the cross section for this process to a relative accuracy of about $20\%$ \cite{CMS:2018uxb, ATLAS:2018mme, CMS:2020cga, ATLAS:2020ior}. With the accumulation of much more data in the near future, the experimental precision is expected to be significantly improved. In order to extract the top Yukawa coupling from the high precision cross section measurements, it is necessary to have high precision theoretical predictions for the relevant observables. 

In quantum chromodynamics (QCD), the total and differential cross sections for $t\bar{t}H$ production at the next-to-leading order (NLO) have been known since almost twenty years ago \cite{Beenakker:2001rj, Reina:2001bc, Reina:2001sf, Beenakker:2002nc, Dawson:2002tg, Dawson:2003zu}. Approximate next-to-next-to-leading order (NNLO) as well as soft gluon resummed results are also calculated in \cite{Kulesza:2015vda, Broggio:2015lya, Broggio:2016lfj, Kulesza:2017ukk, Ju:2019lwp, Broggio:2019ewu, Kulesza:2020nfh, vanBeekveld:2020cat}. These approximations are only valid in certain kinematic limits, and it is highly desirable to have a complete NNLO calculation for this process, in order to control the theoretical uncertainties. However, such a calculation is still out-of-reach due to the obvious obstacles from the complicated two-loop amplitudes. A part of the NNLO contributions that do not involve two-loop amplitudes is recently available \cite{Catani:2021cbl}.

A prominent property of gauge theory amplitudes is the existence of infrared (IR) singularities. Understanding their structure is crucial for designing IR safe quantities that can be compared to experimental measurements. Even in IR safe observables, in certain kinematic regions, there can be large logarithms which need to be resummed to all orders in perturbation theory. The structure of these logarithms is, again, governed by the IR behaviors of scattering amplitudes. In the past couple of years, significant progress has been achieved in the understanding of the IR singularities in non-abelian gauge theories, both in massless \cite{Catani:1998bh, Sterman:2002qn, Aybat:2006wq, Aybat:2006mz, Becher:2009cu, Gardi:2009qi, Dixon:2009gx, Becher:2009qa, Gardi:2009zv, DelDuca:2011ae, Ahrens:2012qz, Naculich:2013xa, Henn:2013wfa, Gardi:2013ita, Falcioni:2014pka, Almelid:2015jia, Almelid:2017qju, Caron-Huot:2017zfo, Becher:2019avh, Agarwal:2020nyc, Agarwal:2021him, Gardi:2021gzz, Falcioni:2021buo, Agarwal:2021ais} and massive \cite{Catani:2000ef, Mitov:2006xs, Becher:2007cu, Czakon:2007ej, Czakon:2007wk, Kidonakis:2009ev, Mitov:2009sv, Becher:2009kw, Ferroglia:2009ep, Ferroglia:2009ii, Mitov:2010xw, Bierenbaum:2011gg, Gardi:2013saa, Vladimirov:2015fea, Kidonakis:2019nqa} cases. Nevertheless, applying these universal behaviors to a given scattering process still requires a considerable amount of work.

In this paper, we apply the method in \cite{Ferroglia:2009ep, Ferroglia:2009ii} to calculate the IR poles in the two-loop amplitudes for the $t\bar{t}H$ production process. The biggest challenge in this calculation is that we need to compute the one-loop amplitude to higher orders in the dimensional regulator $\epsilon=(4-d)/2$, where $d$ is the dimension of spacetime. Unlike the $t\bar{t}$ case, the one-loop $t\bar{t}H$ amplitudes involve many 4-point and 5-point integrals whose higher-order coefficients in $\epsilon$ are not known in the literature. Hence, a major part of this paper is devoted to the systematic calculation of these integrals. These integrals at higher orders in $\epsilon$ are also required for the finite part of the NNLO cross sections.

This paper is organized as follows. In Section~\ref{sec:IRttH} we introduce our notations and review the generic structure of IR singularities of two-loop scattering amplitudes in non-abelian gauge theories. In Section~\ref{sec:OneLoop} we show the details of the calculation of the one-loop amplitudes. In particular, we demonstrate how to construct canonical differential equations for the master integrals. In Section~\ref{sec:res}, we give our results for the IR poles at several representative phase-space points, and briefly summarize our work. We leave some lengthy expressions in the Appendix.

\section{Notations and structure of IR singularities}
\label{sec:IRttH}

For the production of a Higgs boson associated with a top-quark pair, we consider the partonic processes
\begin{align}
q_\beta(p_1)+\bar q_\alpha(p_2) &\to t_k(p_3)+\bar t_l(p_4)+H(p_5) \,, \nonumber \\
g_a(p_1)+g_b(p_2) &\to t_k(p_3)+\bar t_l(p_4)+H(p_5) \,,
\end{align}
where $\alpha,\beta,k,l,a,b$ are color indices.
We define the following kinematic variables:
\begin{gather}
s_{ij} \equiv (\sigma_i p_i + \sigma_j p_j)^2 \, , \quad \tilde{s}_{ij}=2\sigma_i\sigma_j p_i\cdot p_j \nonumber \\ p_1^2=p_2^2=0 \, ,\quad p_3^2 = p_4^2 = m_t^2 ,\  \quad p_5^2 = m_H^2 \, ,
\end{gather}
where $\sigma_i=+1$ if $p_i$ is incoming, and $\sigma_i=-1$ if $p_i$ is outgoing. We use the color space formalism \cite{Catani:1996jh,Catani:1996vz} where the amplitudes are vectors $\ket{{\cal M}_{q,g}}$. The subscript $q$ or $g$ specifies the quark-antiquark annihilation channel or the gluon fusion channel, respectively. For the $q\bar{q}$ channel, we choose the independent color structures as
\begin{equation}\label{eq:qqbasis}
   \ket{c_1} = \delta_{\alpha\beta}\,\delta_{kl} \,, 
    \quad 
   \ket{c_2} = (t^a)_{\alpha\beta}\,(t^a)_{kl} \,.
\end{equation}
For the $gg$ channel, we use the color basis
\begin{equation}
   \ket{c_1} = \delta^{ab}\,\delta_{kl} \,,
    \quad 
   \ket{c_2} = if^{abc}\,(t^c)_{kl} \,,
    \quad 
   \ket{c_3} = d^{abc}\,(t^c)_{kl} \,.
\end{equation}
The UV divergences in the amplitudes are renormalized according to 
\begin{align}
\Ket{{\cal M}^R_{q,g}(\alpha_s, g_Y, m_t, \mu, \epsilon)} = \left(\frac{\mu^2 e^{\gamma_E}}{4\pi}\right)^{-3\epsilon/2} Z_{q,g} Z_Q \Ket{{\cal M}^{\text{bare}}_{q,g}(\alpha_s^0, g_Y^0, m_t^0,\epsilon)} ,
\label{eq:UVren}
\end{align}
where $Z_g$, $Z_q$ and $Z_Q$ are the on-shell wave-function renormalization constants for gluons, light- and heavy-quarks, respectively. We have suppressed the dependence of the amplitudes on other kinematic variables. The Yukawa coupling $g_Y$ is defined as
\begin{equation}
g_Y = \frac{e \, m_t}{2m_W\sin(\theta_W)} \, .
\end{equation}
We renormalize the top quark mass in the on-shell scheme: $m_t^0 = Z_m m_t$, and the Yukawa coupling is renormalized accordingly. The strong coupling constant $\alpha_s$ is renormalized in the $\overline{\text{MS}}$ scheme with $n_f=n_l+n_h$ active flavors. The relations between the bare couplings and the renormalized ones are given by
\begin{equation}
\alpha_s^0 = \left( \frac{\mu^2 e^{\gamma_E}}{4\pi} \right)^{\epsilon} Z_{\alpha_s} \alpha_s \, , \quad g_Y^0 = \left( \frac{\mu^2 e^{\gamma_E}}{4\pi} \right)^{\epsilon/2} Z_m \, g_Y \, .
\end{equation}
The renormalization constants are
\begin{align}
Z_{q} &= 1 + \mathcal{O}(\alpha_s^2) \, , \nonumber
\\
Z_{g} &= 1 + \frac{\alpha_s}{4\pi} T_Fn_h \left[ -\frac{4}{3\epsilon} - \frac{4}{3}L_{\mu} - \epsilon\left(\frac{\pi^2}{9} + \frac{2}{3}L^2_{\mu} \right)\right] + \mathcal{O}(\alpha_s^2) \, , \nonumber
\\
Z_{Q} &= 1 + \frac{\alpha_s}{4\pi} C_F \left[ -\frac{3}{\epsilon} - \left(4+3L_\mu\right) - \epsilon\left(8+4L_\mu+\frac{\pi^2}{4}+\frac{3}{2}L_\mu^2\right) \right] + \mathcal{O}(\alpha_s^2) \, , \nonumber
\\
Z_{\alpha_s} &= 1 - \frac{\alpha_s}{4\pi} \frac{\beta_0}{\epsilon} +\mathcal{O}(\alpha_s^2) \, , \nonumber \\ 
Z_m &= 1 + \frac{\alpha_s}{4\pi} C_F \left[ -\frac{3}{\epsilon} - \left(4+3L_\mu\right) - \epsilon\left(8+4L_\mu+\frac{\pi^2}{4}+\frac{3}{2}L_\mu^2\right) \right] + \mathcal{O}(\alpha_s^2) \, , 
\end{align}
where $L_\mu=\ln(\mu^2/m_t^2)$, $T_F = 1/2$, $C_F = (N^2-1)/2N$, $\beta_0 = 11N/3 - 4T_Fn_f/3$, with $N=3$ being the number of colors.

After UV renormalization, the remaining IR divergences can be subtracted by a multiplicative factor $\bm{Z}^{-1}(\epsilon, m_t, \mu)$, where the bold symbol denotes an operator in color space. More precisely, we have
\begin{align}
\label{eq:key}
\bm{Z}_{q,g}^{-1}(\epsilon, m_t, \mu) \Ket{{\cal M}^R_{q,g}(\alpha_s, g_Y, m_t, \mu, \epsilon)} = \text{finite} \,.
\end{align}
The $\bm{Z}$ factor satisfies a renormalization group equation (RGE) of the form
\begin{align}
\bm{Z}^{-1}(\epsilon, m_t, \mu) \frac{d}{d\log\mu} \bm{Z}(\epsilon, m_t, \mu) = -\bm{\Gamma}(m_t, \mu) \, ,
\end{align}
where $\bm{\Gamma}$ is a universal anomalous-dimension operator, which has been calculated up to order $\alpha_s^2$ in \cite{Becher:2009kw,Ferroglia:2009ep,Ferroglia:2009ii}.

In the $t\bar{t}H$ production process, the anomalous-dimension matrices for the $q\bar{q}$ and $gg$ amplitudes are rather similar as those in the $t\bar{t}$ production process considered in \cite{Ferroglia:2009ii,Ahrens:2010zv}. They are given by \cite{Broggio:2015lya}
\begin{align}
   \bm{\Gamma}_{q\bar q} 
   &= \left[ C_F\,\gamma_{\rm cusp}(\alpha_s)\,\log\frac{-s_{12}}{\mu^2}
    + C_F\,\gamma_{\rm cusp}(\beta_{34},\alpha_s)
    + 2\gamma^q(\alpha_s) + 2\gamma^Q(\alpha_s) \right] \bm{1} 
\nonumber 
\\
   &\quad + \frac{N}{2} \left[ 
    \gamma_{\rm cusp}(\alpha_s)\,
    \log\frac{(-\tilde{s}_{13})(-\tilde{s}_{24})}{(-s_{12})\,m_t^2}
    - \gamma_{\rm cusp}(\beta_{34},\alpha_s) \right]
    \begin{pmatrix}
     0~ & ~0 \\ 0~ & ~1
    \end{pmatrix}
\nonumber
\\
   &\quad + \gamma_{\rm cusp}(\alpha_s)\,
    \log\frac{(-\tilde{s}_{13})(-\tilde{s}_{24})}{(-\tilde{s}_{14})(-\tilde{s}_{23})} \left[
    \begin{pmatrix}
     0~ & \frac{C_F}{2N} \\
     1~ & - \frac{1}{N}
    \end{pmatrix} 
    + \frac{\alpha_s}{4\pi}\,g(\beta_{34})
    \begin{pmatrix}
     0 & \frac{C_F}{2} \\ -N & 0
    \end{pmatrix} \right]
    + {\cal O}(\alpha_s^3) \,,
\nonumber
\\
   \bm{\Gamma}_{gg} 
   &= \left[ N\,\gamma_{\rm cusp}(\alpha_s)\,\log\frac{-s_{12}}{\mu^2} 
    + C_F\,\gamma_{\rm cusp}(\beta_{34},\alpha_s)
    + 2\gamma^g(\alpha_s) + 2\gamma^Q(\alpha_s) \right] \bm{1} 
    \nonumber\\
   &+ \frac{N}{2} \left[ 
    \gamma_{\rm cusp}(\alpha_s)\,
    \log\frac{(-\tilde{s}_{13})(-\tilde{s}_{24})}{(-s_{12})\,m_t^2}
    - \gamma_{\rm cusp}(\beta_{34},\alpha_s) \right]
    \begin{pmatrix}
     0~ & ~0~ & ~0 \\
     0~ & ~1~ & ~0 \\
     0~ & ~0~ & ~1
    \end{pmatrix}
\nonumber
\\
   &+ \gamma_{\rm cusp}(\alpha_s)\,
    \log\frac{(-\tilde{s}_{13})(-\tilde{s}_{24})}{(-\tilde{s}_{14})(-\tilde{s}_{23})} \left[
    \begin{pmatrix}
     0~ & \frac{1}{2} & 0 \\
     1~ & - \frac{N}{4} & \frac{N^2-4}{4N} \\
     0~ & \frac{N}{4} & - \frac{N}{4}
    \end{pmatrix} 
    + \frac{\alpha_s}{4\pi}\,g(\beta_{34})
    \begin{pmatrix}
     0 & \frac{N}{2} & 0~ \\ -N & ~0~ & 0~ \\ 0 & ~0~ & 0~
    \end{pmatrix} \right] 
    + {\cal O}(\alpha_s^3) \,.
\end{align}
where 
\begin{equation}
   g(\beta) = \coth\beta \left[ \beta^2 
    + 2\beta\,\log(1-e^{-2\beta}) - \mbox{Li}_2(e^{-2\beta}) 
    + \frac{\pi^2}{6} \right] - \beta^2 - \frac{\pi^2}{6} \, .
\end{equation}
The cusp angle $\beta_{34}$ is defined by
\begin{align}
\cosh \beta_{34} = -\frac{\tilde{s}_{34}}{2m_t^2} \,.
\end{align}
The perturbative expansions of $\gamma_{\text{cusp}}$, $\gamma^q$, $\gamma^g$ and $\gamma^Q$ can be found, for instance, in the Appendix of \cite{Ahrens:2010zv}. The only difference of the anomalous-dimension matrices here (with respect to those in $t\bar{t}$ production) is that $\tilde{s}_{13} \neq \tilde{s}_{24}$ and $\tilde{s}_{14} \neq \tilde{s}_{23}$ due to the $2 \to 3$ kinematics.

Both the UV-renormalized amplitudes and the IR subtraction factors can be expanded in powers of $\alpha_s$:
\begin{align}
   \Ket{{\cal M}_{q,g}^R} &= \frac{4\pi\alpha_s g_Y}{m_t}  \left[ \Ket{{\cal M}_{q,g}^{(0)}} 
    + \frac{\alpha_s}{4\pi} \Ket{{\cal M}_{q,g}^{(1)}} 
    + \left( \frac{\alpha_s}{4\pi} \right)^2 \Ket{{\cal M}_{q,g}^{(2)}} 
    + \cdots \right] , \nonumber
    \\
    \bm{Z}_{q,g} &= \bm{1} + \frac{\alpha_s}{4\pi} \bm{Z}_{q,g}^{(1)} + \left( \frac{\alpha_s}{4\pi} \right)^2 \bm{Z}_{q,g}^{(2)} + \cdots \, .
\end{align}
We may then extract the IR singularities of the amplitudes order-by-order in $\alpha_s$:
\begin{align}
\label{eq:IRamps}
   \Ket{{\cal M}_{q,g}^{(1),\,\text{sing}}} 
   &= \bm{Z}_{q,g}^{(1)} \Ket{{\cal M}_{q,g}^{(0)}} , \nonumber
   \\
   \Ket{{\cal M}_{q,g}^{(2),\,\text{sing}}} 
   &= \left[ \bm{Z}_{q,g}^{(2)} - \left( \bm{Z}_{q,g}^{(1)}\right)^2 \right]
    \Ket{{\cal M}_{q,g}^{(0)}}
    + \left( \bm{Z}_{q,g}^{(1)} \Ket{{\cal M}_{q,g}^{(1)}} \right)_{\text{poles}} .
\end{align}
Note that to predict the IR poles at the two-loop order, one must calculate the UV-renormalized one-loop amplitudes to ${\cal O}(\epsilon^1)$. They multiply the $1/\epsilon^2$ terms in $\bm{Z}_{q,g}^{(1)}$ and give rise to $1/\epsilon$ divergences. The next section is devoted to this non-trivial task.

\section{Calculation of the one-loop amplitudes to higher orders in $\epsilon$}
\label{sec:OneLoop}

\subsection{Setup}

As is clear from the last section, in order to predict the IR poles at two loops, we need the one-loop amplitudes up to order $\epsilon^1$. We generate the amplitudes using \texttt{FeynArts} \cite{Hahn:2000kx}, and manipulate the expressions with \texttt{FeynCalc} \cite{Shtabovenko:2020gxv}. We then need to express the amplitudes in terms of scalar integrals, there are two ways to achieve this. Since we are interested in the IR poles in the interference of $\mathcal{M}^{(2)}$ with $\mathcal{M}^{(0)}$, we may readily multiply the one-loop amplitudes by the tree-level ones. The Lorentz contractions and Dirac traces can now be easily performed while keeping the color information. Alternatively, we may also apply a complete set of projectors to the one-loop amplitudes, and extract the coefficients as a linear combination of scalar integrals.\footnote{The projectors are similar to those in $t\bar{t}$ production and can be found in \cite{Beenakker:2002nc}.} The second method is more complicated for our purpose, but the results can be useful if one wants to obtain the one-loop amplitude squared. We have performed the calculation in both ways, and the results agree.

\begin{figure}[t!]
\centering
\begin{minipage}[t]{0.4\textwidth}
\centering
\includegraphics[scale=0.6]{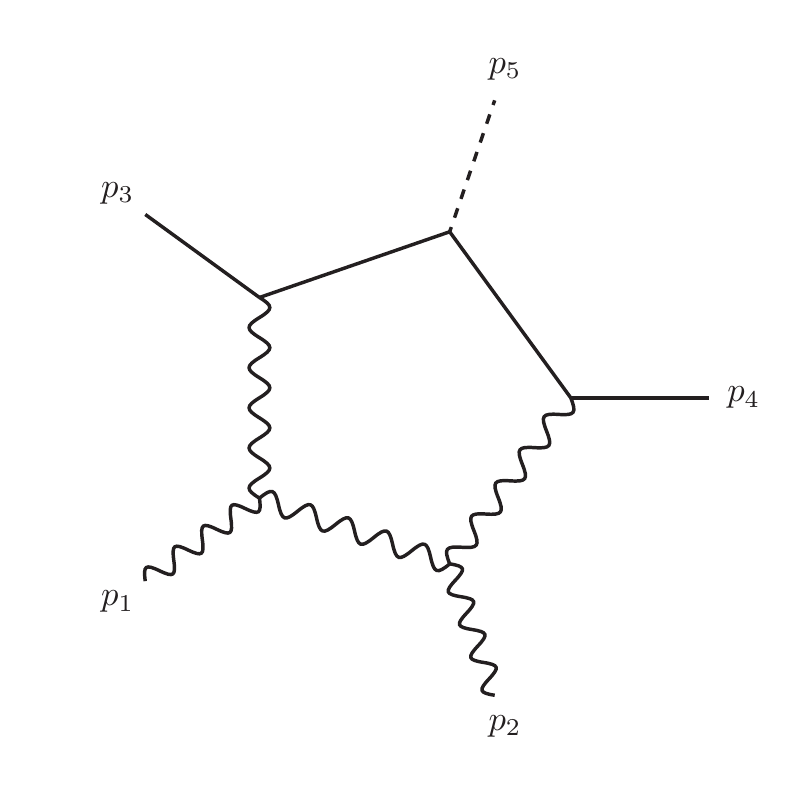}
\\
{\small Topology A}
\end{minipage}
\begin{minipage}[t]{0.4\textwidth}
\centering
\includegraphics[scale=0.6]{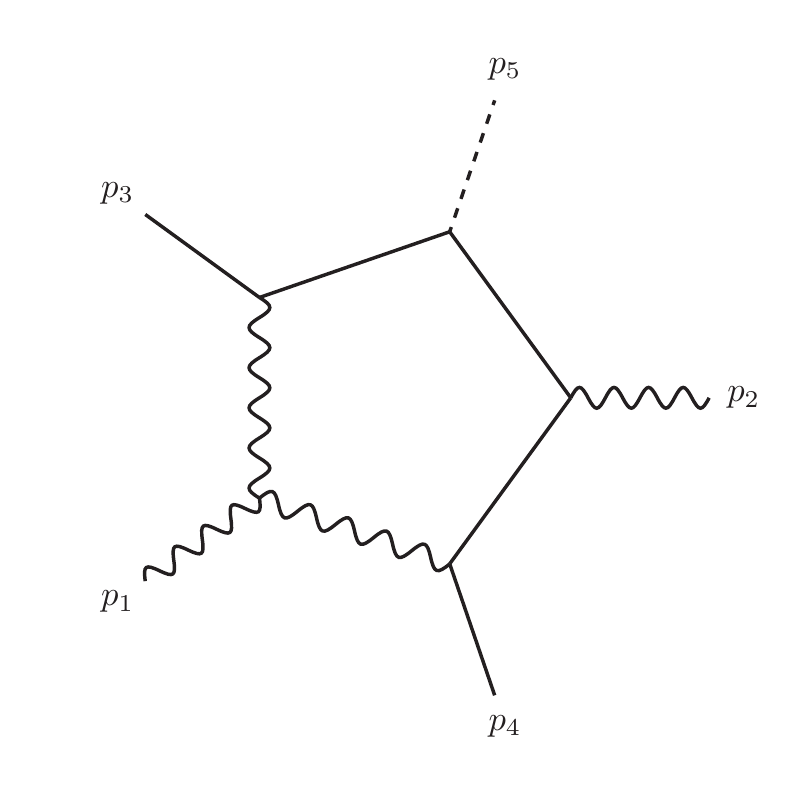}
\\
{\small Topology B}
\end{minipage}
\\
\begin{minipage}[t]{0.4\textwidth}
\centering
\includegraphics[scale=0.6]{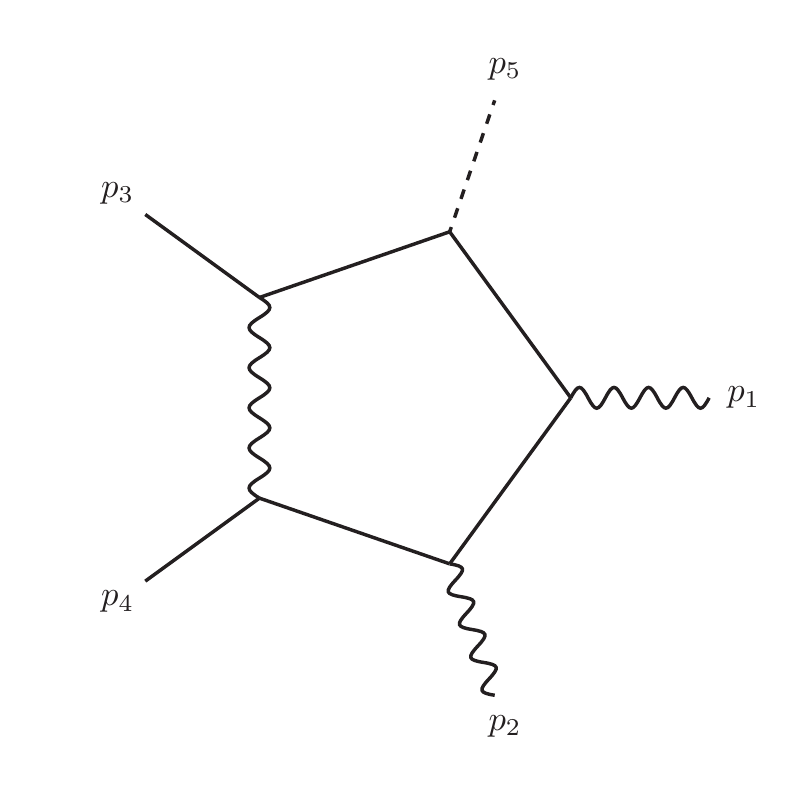}
\\
{\small Topology C}
\end{minipage}
\begin{minipage}[t]{0.4\textwidth}
\centering
\includegraphics[scale=0.6]{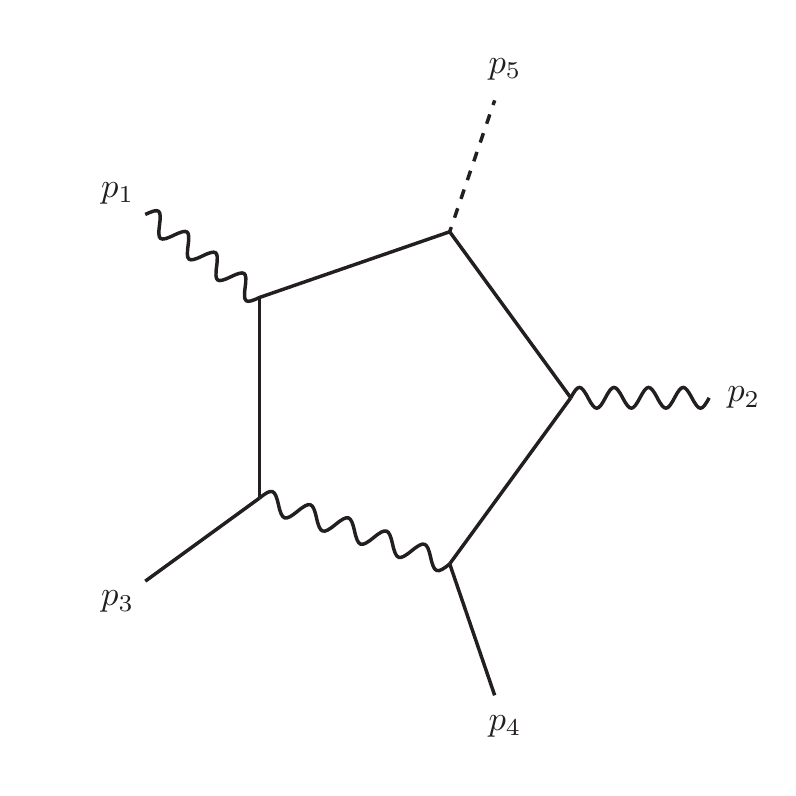}
\\
{\small Topology D}
\end{minipage}
\caption{\label{fig:topo}The four independent integral families. Wiggly lines are massless; solid lines have mass $m_t$; and dashed lines have mass $m_H$.}
\end{figure}

The one-loop scalar integrals can be categorized into 12 families (topologies), four of which are independent (and the others can be obtained with exchanges of external momenta). Each family is defined by 5 propagator denominators denoted as $D_i\, (i=1,\ldots,5)$:
\begin{equation}
\label{eq:def_integral}
F_{a_1,a_2,a_3,a_4,a_5} = \frac{  (m_t^2)^{a-2+\epsilon}}{ \Gamma(\epsilon)} \int \frac{d^dl}{i \pi^{d/2}} \frac{1}{D_1^{a_1} D_2^{a_2} D_3^{a_3} D_4^{a_4} D_5^{a_5}} \, , 
\end{equation}
where $a \equiv a_1+a_2+a_3+a_4+a_5$. The prefactor is introduced such that the integrals are dimensionless and do not contain $\gamma_E$ in their series expansions in $\epsilon$. The propagator denominators for the four independent families are chosen as
\begin{align}\label{eq:topo}
&T_A=\{l^2-m_t^2,\  (l-p_3)^2,\ (l+p_1-p_3)^2,\ (l+p_1+p_2-p_3)^2,\ \nonumber \\  
&\qquad\quad\ (l+p_1+p_2-p_3-p_4)^2-m_t^2\} \,, \nonumber \\
&T_B=\{l^2-m_t^2, \ (l-p_3)^2, \ (l+p_1-p_3)^2, \ (l+p_1-p_3-p_4)^2-m_t^2, \ \nonumber\\  
&\qquad\quad\  (l+p_1+p_2-p_3-p_4)^2-m_t^2\} \,, \nonumber  \\
&T_C=\{l^2-m_t^2, \ (l-p_3)^2, \ (l-p_3-p_4)^2-m_t^2, \ (l+p_2-p_3-p_4)^2-m_t^2,\ \nonumber\\  
&\qquad\quad\ (l+p_1+p_2-p_3-p_4)^2-m_t^2\} \,, \nonumber \\
&T_D=\{l^2-m_t^2, \ (l+p_1)^2-m_t^2, \ (l+p_1-p_3)^2, \ (l+p_1-p_3-p_4)^2-m_t^2, \ \nonumber\\  
&\qquad\quad\ (l+p_1+p_2-p_3-p_4)^2-m_t^2\} \,.
\end{align}
The corresponding diagrams are depicted in Figure~\ref{fig:topo}. The remaining 8 topologies can be obtained by the exchanges $p_1 \leftrightarrow p_2$ and/or $p_3 \leftrightarrow p_4$. There are 18, 20, 22 and 22 master integrals in the topologies A, B, C and D, respectively.

To calculate the master integrals, we adopt the method of canonical differential equations \cite{Henn:2013pwa}. Namely, we construct linear combinations of the master integrals which satisfy a set of differential equations of the $\epsilon$-form.  We denote such a ``canonical basis'' as $\vec{f}$. Integrals in such a basis have the property of uniform transcendentality (UT), and hence are also dubbed ``UT integrals''. We introduce the dimensionless kinematic variables
\begin{align}
x_{ij} = \frac{s_{ij}}{m_t^2} \,, \quad x_h = \frac{m_H^2}{m_t^2} \, .
\end{align}
The differential equations can then be written as
\begin{equation}
d\vec{f}(\vec{x}, \epsilon) = \epsilon \, d\bm{A}(\vec{x}) \, \vec{f}(\vec{x}, \epsilon) \, ,
\label{eq:can_de}
\end{equation}
where $\vec{x}$ denotes a set of independent kinematic variables chosen from $x_{ij}$ and $x_h$ (note that the choices of independent variables are different for each topology). The matrix $d\bm{A}$ takes the $d\log$-form:
\begin{equation}
d\bm{A}(\vec{x}) = \sum_i \bm{C}_i \, d\log(W_i(\vec{x})) \, ,
\label{eq:dA_Wi}
\end{equation}
where $\bm{C}_i$ are matrices consisting of rational numbers, and $W_i(\vec{x})$ are algebraic functions of the kinematic variables. The functions $W_i$ are called the ``letters'' for this topology, and the set of all independent letters is called the ``alphabet''.

The canonical differential equations \eqref{eq:can_de} can be solved order-by-order in $\epsilon$. To this end, we expand the (suitably normalized) integrals as Taylor series
\begin{equation}
\vec{f}(\vec{x},\epsilon) = \sum_{n=0}^{\infty} \epsilon^n \, \vec{f}^{(n)}(\vec{x}) \, ,
\end{equation}
where the $n$th order coefficient function can be written as a Chen iterated integral \cite{Chen:1977oja}
\begin{equation}
\vec{f}^{(n)}(\vec{x}) = \int_{\vec{x}_0}^{\vec{x}} d\bm{A}(\vec{x}_n) \cdots \int_{\vec{x}_0}^{\vec{x}_3} d\bm{A}(\vec{x}_2) \int_{\vec{x}_0}^{\vec{x}_2} d\bm{A}(\vec{x}_1) + \vec{f}^{(n)}(\vec{x}_0) \, .
\end{equation}
In certain cases, these iterated integrals can be solved analytically (either by direct integration or by bootstrapping). The results can often be written in terms of generalized polylogarithms (GPLs) which allow efficient numeric evaluation \cite{Vollinga:2004sn, Naterop:2019xaf, Wang:2021imw}. When an analytic solution is not available, it is straightforward to evaluate them numerically either by numerical integration or by a series expansion \cite{Moriello:2019yhu, Hidding:2020ytt}.
In the rest of this Section, we discuss the construction of the canonical basis $\vec{f}$ and the matrix $d\bm{A}(\vec{x})$.

\subsection{The canonical master integrals}

We use the method of \cite{Chen:2020uyk} to construct the canonical bases using the Baikov representation~\cite{Baikov:1996iu}. We present the results with generic external momenta and internal masses. The results for the $t\bar{t}H$ process can be obtained by inserting the momenta and masses associated with the propagator denominators for each topology.

Consider a generic one-loop integral family with $N = E + 1$ external legs, where $E$ is the number of independent external momenta. Integrals in this family can be written as
\begin{align}\label{eq:intI}
I_{a_1, \cdots, a_N} = \int \frac{d^dk}{i \pi^{d/2}} \frac{1}{z_1^{a_1} z_2^{a_2}\cdots z_N^{a_N}} \,,
\end{align}
where $z_i$ are the propagator denominators given by
\begin{align}
\label{eq:propagator_general}
z_1=k^2-m_1^2 \,, \quad z_2= (k+q_1)^2-m_2^2 \,, \cdots \,, \quad z_N= (k+q_1+\cdots+q_{E})^2-m_N^2 \,.
\end{align}
The corresponding Baikov representation is given by
\begin{align}
I_{a_1,\ldots,a_N} &= \frac{e^{\epsilon\gamma_E}}{(4 \pi)^{E/2} \, \Gamma\big((d-E)/2\big)} \int_{\mathcal{C}} \frac{ \left| G_N(\bm{z}) \right|^{(d-E-2)/2}}{\left| \mathcal{K}_N \right|^{(d-E-1)/2}  }  \prod_{i=1}^{N} \frac{dz_i}{z_i^{a_i}} \, ,
\label{eq:1loopbaikov}
\end{align}
where $\bm{z}=\{z_1,\ldots,z_N\}$ is the collection of the Baikov variables (i.e., propagator denominators). The function $G_N(\bm{z})$ is a polynomial of the $N$ variables, while $\mathcal{K}_N$ is independent of $\bm{z}$. They are given by
\begin{equation}
G_N(\bm{z}) \equiv G(k,q_1,\ldots,q_E) \, , \quad \mathcal{K}_N = G(q_1, \cdots, q_{E}) \, ,
\label{eq:GN}
\end{equation}
where the Gram determinant is defined as
\begin{equation}
\label{eq:gram1}
G(q_1,\ldots,q_n) \equiv  \det
\begin{pmatrix}
q_1 \cdot q_1 & q_1 \cdot q_2 & \cdots & q_1 \cdot q_n
\\
q_2 \cdot q_1 & q_2 \cdot q_2 & & \vdots
\\
\vdots & & \ddots & \vdots
\\
q_n \cdot q_1 & \cdots & \cdots & q_n \cdot q_n
\end{pmatrix}
\, .
\end{equation}

The UT integrals $g_N$ for any $N$ is obtained in \cite{Chen:2020uyk}. For the purpose of this work, we need them up to $N = 5$. They are given by
\begin{align}
\label{dlog_nPoint}
g_1 &= \frac{\epsilon}{\Gamma(1-\epsilon)} \int \left(\frac{1}{G_1(z_1)}\right)^\epsilon \frac{dz_1}{z_1} \,,\nonumber  \\
g_2 &= \frac{\epsilon}{2\sqrt{\pi}\Gamma(1/2-\epsilon)} \int \frac{\sqrt{G_2(0,0)}}{\sqrt{G_2(z_1,z_2)}}\left(-\frac{\mK_2}{G_2(z_1,z_2)}\right)^\epsilon \frac{dz_1}{z_1}\frac{dz_2}{z_2} \,, \nonumber \\
g_3 &= \frac{\epsilon^2}{4\pi\Gamma(1-\epsilon)} \int \left(-\frac{\mK_3}{G_3(z_1,z_2,z_3)} \right)^{\epsilon}\frac{dz_1}{z_1}\frac{dz_2}{z_2}\frac{dz_3}{z_3} \,, \nonumber\\
g_4 &= \frac{\epsilon^2}{8\pi^{3/2}\Gamma(1/2-\epsilon)} \int \frac{\sqrt{G_4(0,0,0,0)}}{\sqrt{G_4(z_1,z_2,z_3,z_4)}}\left(-\frac{\mK_4}{G_4(z_1,z_2,z_3,z_4)}\right)^\epsilon \frac{dz_1}{z_1}\frac{dz_2}{z_2}\frac{dz_3}{z_3}\frac{dz_4}{z_4} \,, \nonumber\\
g_5 &= \frac{\epsilon^3}{16\pi^2\Gamma(1-\epsilon)} \int \left(-\frac{\mK_5}{G_5(z_1,z_2,z_3,z_4,z_5)}\right)^\epsilon \frac{dz_1}{z_1}\frac{dz_2}{z_2}\frac{dz_3}{z_3}\frac{dz_4}{z_4}\frac{dz_5}{z_5} \,.
\end{align}
Note that $g_3$ and $g_4$ can be straightforwardly identified as Feynman integrals in $4-2\epsilon$ dimensions. On the other hand, $g_1$ and $g_2$ can be naturally regarded as creatures in $2-2\epsilon$ dimensions, while $g_5$ lives in $6-2\epsilon$ dimensions. Namely we have:
\begin{align}
g_1 &=\epsilon \, I_{1}^{(2-2\epsilon)} \,, \nn\\
g_2 &=\epsilon \sqrt{G_2(0,0)} \, I_{1,1}^{(2-2\epsilon)} \,, \nn\\
g_3 &=\epsilon^2 \sqrt{\mK_3} \, I_{1,1,1}^{(4-2\epsilon)} \,, \nn\\
g_4 &=\epsilon^2 \sqrt{G_4(0,0,0,0)} \, I_{1,1,1,1}^{(4-2\epsilon)} \,, \nn\\
g_5 &=\epsilon^3 \sqrt{\mK_5} \, I_{1,1,1,1,1}^{(6-2\epsilon)} \,. 
\label{eq:UTMIs}
\end{align}
The $2-2\epsilon$ and $6-2\epsilon$ dimensional integrals can be expressed as Feynman integrals in $4-2\epsilon$ dimensions using the dimensional recurrence relations \cite{Tarasov:1996br, Lee:2009dh}. Applying the above to all sectors of a family, we build a complete canonical basis satisfying $\epsilon$-form differential equations. As a final remark, we note that there is a freedom in multiplying a UT integral by some complex number, and the result remains UT. We will use this freedom when writing down the canonical basis for $t\bar{t}H$ production listed in the Appendix.

\subsection{Bootstrapping the coefficient matrices in the differential equations}
\label{sec:letter}

Given the UT integrals in \eqref{eq:UTMIs}, it is straightforward to calculate their derivatives with respect to some kinematic variable $x_i$:
\begin{equation}
\frac{\partial}{\partial x_i} \vec{f}(\vec{x},\epsilon) = \epsilon \, \bm{A}_i(\vec{x}) \, \vec{f}(\vec{x},\epsilon) \, ,
\end{equation}
where the elements in the matrix $\bm{A}_i(\vec{x})$ have the property that they only contain simple poles. We would like to combine these derivatives into a total derivative as in Eq.~\eqref{eq:can_de}. We will achieve this by bootstrapping. According to Eq.~\eqref{eq:dA_Wi}, we can write the total derivative as
\begin{equation}
d\vec{f}(\vec{x},\epsilon) = \sum_i \frac{\partial}{\partial x_i} \vec{f}(\vec{x},\epsilon) \, dx_i = \epsilon \sum_j \bm{C}_j \, d\log(W_j(\vec{x})) \vec{f}(\vec{x},\epsilon) \, .
\end{equation}
This leads to
\begin{equation}
\sum_j \bm{C}_j \, \frac{\partial \log(W_j(\vec{x}))}{x_i} = \bm{A}_i(\vec{x}) \, , \quad \forall i \, .
\end{equation}
Since $\bm{A}_i(\vec{x})$ are known, it is easy to extract the coefficient matrices $\bm{C}_j$ once we know the letters $W_j(\vec{x})$.

We obtain the full alphabet using the method described in \cite{Abreu:2017ptx, Abreu:2017enx, Abreu:2017mtm, Chen:2022fyw}. We write the differential equation satisfied by an $N$-point UT integral $g_N$ (see Eqs.~\eqref{dlog_nPoint} and \eqref{eq:UTMIs}) as
\begin{align}
dg_N(\vec{x},\epsilon) = \epsilon \, dM_N(\vec{x}) \, g_N(\vec{x},\epsilon) + \epsilon \sum_{m < N} \sum_i  dM_{N,m}^{(i)}(\vec{x}) \,   g_m^{(i)}(\vec{x},\epsilon) \, ,
\end{align}
where $g_N(\vec{x},\epsilon)$ and $g_m^{(i)}(\vec{x},\epsilon)$ are components of the canonical basis $\vec{f}(\vec{x},\epsilon)$, with the superscript $(i)$ labelling different $m$-point integrals. The entries $dM_N(\vec{x})$ and $dM_{N,m}^{(i)}(\vec{x})$
belong to the matrix $d\bm{A}(\vec{x})$, and can be written in the form
\begin{equation}
dM_N(\vec{x}) \propto d\log(W_N(\vec{x})) \, , \quad dM_{N,m}^{(i)}(\vec{x}) \propto d\log(W_{N,m}^{(i)}(\vec{x})) \, .
\end{equation}
In the following, we present the generic form of the letters $W_N$ and $W_{N,m}^{(i)}$ obtained in \cite{Chen:2022fyw}. For each $m$ we take $g_m$ to be the UT integral with the denominators $z_1,\ldots,z_m$, and show the corresponding $W_{N,m}$. The other ones can be obtained by rearranging the order of denominators.

The self-dependent letter $W_N$ is given by
\begin{equation}
W_N = \frac{\widetilde{G}_N}{\mK_N} \, ,
\end{equation}
where $\widetilde{G}_N \equiv G_N(\bm{0})$. As an example, we consider a 5-point integral in the topology A of $t\bar{t}H$ production. The Gram determinants entering the letter $W_5$ are given by:
\begin{align}
\mK_5 &= G(-p_3, p_1, p_2, -p_4)  \nonumber \\
&= \frac{1}{16} \Big( x_{12}^2 x_h^2-2 x_{13} x_{12}^2 x_h-2 x_{24} x_{12}^2 x_h+2 x_{13} x_{12} x_h-4 x_{13} x_{24} x_{12} x_h+2
   x_{24} x_{12} x_h     \nonumber  \\
   &+2 x_{12} x_{35} x_{13} x_h-2 x_{12} x_h+2 x_{12} x_{24} x_{45} x_h-2 x_{12} x_{35} x_{45} x_h+x_{12}^2
   x_{13}^2+x_{12}^2 x_{24}^2  \nonumber \\
   &-2 x_{13} x_{24} x_{12}^2-2 x_{13}^2 x_{12}-2 x_{24}^2 x_{12}+2 x_{13} x_{12}+4 x_{13} x_{24} x_{12}-2 x_{13}^2 x_{35} x_{12} \nonumber \\
   &-2 x_{12} x_{45} x_{24}^2+2 x_{12} x_{13} x_{35} x_{24}-4 x_{12} x_{45} x_{24}+2 x_{12} x_{13} x_{45}
   x_{24}-4 x_{12} x_{13} x_{35}  \nonumber \\
   &+x_{35}^2 x_{13}^2+x_{13}^2+2 x_{12} x_{35} x_{45} x_{13}-2 x_{13}+x_{24}^2 x_{45}^2+x_{35}^2
   x_{45}^2+2 x_{12} x_{24} x_{35} x_{45} \nonumber \\
   &-2 x_{35} x_{13}^2+2 x_{24} x_{13}-2 x_{24} x_{35} x_{13}+2 x_{35} x_{13}-2 x_{24} x_{35} x_{45}^2-2
   x_{24}-2 x_{24}^2 x_{45}  \nonumber\\
   &-2 x_{13} x_{45} x_{35}^2+2 x_{13} x_{45} x_{35}+2 x_{13} x_{24} x_{45} x_{35}-2 x_{13} x_{24} x_{45}+2
   x_{24} x_{45} \nonumber \\
   &+2x_{24} x_{35} x_{45}-2 x_{35} x_{45}+2 x_{24} x_{12} +x_{24}^2 +1  \Big) \,, \nonumber \\
\widetilde{G}_5 &= G(l,-p_3 ,p_1, p_2, -p_4)\big|_{\bm{z}=\bm{0}}  \nonumber \\
&= \frac{1}{16}x_{12} \Big(x_{12} x_{24} x_{13} x_h -x_{12} x_{13} x_h+x_{12} x_h-x_{12} x_{24} x_h+x_{12} x_{13}^2-2 x_{12} x_{24} x_{13} +x_{45}\nonumber \\
&-x_{13} x_{24}+x_{13} x_{35} x_{24}+x_{13} x_{45} x_{24}-x_{13} x_{35} x_{45} x_{24}-x_{13} x_{35}+x_{13}
   x_{35} x_{45} +x_{35}\nonumber \\
&+x_{12} x_{24}^2-x_{35} x_{24}+x_{35} x_{45} x_{24}-x_{45} x_{24}+x_{24}+x_{13}-x_{13} x_{45}-x_{35}
   x_{45}-1\Big) \,.
\end{align}
Here we have omitted the dimension of Gram determinants since they only appear in dimensionless letters.

The letter $W_{N,N-1}$ with odd $N$ is given by
\begin{equation}
W_{N,N-1} = \frac{\widetilde{B}_{N}-\sqrt{\widetilde{G}_{N-1}\mathcal{K}_N}}{\widetilde{B}_{N}+\sqrt{\widetilde{G}_{N-1}\mathcal{K}_N}} \, ,
\label{eq:W_n1_odd}
\end{equation}
where $\widetilde{B}_N \equiv B_N(\bm{0})$ with
\begin{equation}
B_N(\bm{z}) = G(k,q_1,\ldots,q_{E-1};\, q_E,q_1,\ldots,q_{E-1}) \, ,
\end{equation}
where the extended Gram determinant is defined as
\begin{align}
G(q_1,\ldots,q_n;\, k_1,\ldots,k_n) = \det
\begin{pmatrix}
q_1 \cdot k_1 & q_1 \cdot k_2 & \cdots & q_1 \cdot k_n
\\
q_2 \cdot k_1 & q_2 \cdot k_2 & & \vdots
\\
\vdots & & \ddots & \vdots
\\
q_n \cdot k_1 & \cdots & \cdots & q_n \cdot k_n
\end{pmatrix}
.
\end{align}
We again show an example in topology A for $t\bar{t}H$ production. Consider the contribution from $F_{1,1,1,1,0}$ to the differential equation of $F_{1,1,1,1,1}$, we need the Gram determinants
\begin{align}
\widetilde{G}_4 &= G(l,-p_3, p_1, p_2)\big|_{\bm{z}=\bm{0}} =\frac{1}{16} x_{12}^2(x_{13}-1)^2 \,, \nonumber\\
\widetilde{B}_5 &= G(l,-p_3, p_1, p_2 ;\, -p_4, -p_3, p_1,p_2)\big|_{\bm{z}=\bm{0}} \nonumber \\
&=-\frac{1}{16}\Big(x_{12}^2 x_h-x_{13} x_{12}^2 x_h+x_{13}^2 x_{12}^2+x_{13} x_{12}^2-x_{13}
   x_{24} x_{12}^2-x_{24} x_{12}^2+x_{13}^2 x_{12} \nonumber \\
   &-x_{12} x_{35} x_{13}^2-x_{12} x_{24} x_{13}+x_{12} x_{35} x_{13}-2 x_{12}
   x_{45} x_{13}+x_{12} x_{24} x_{45} x_{13}+x_{12} x_{24} \nonumber \\
   &-x_{24} x_{45} x_{12}+x_{13} x_{35} x_{45} x_{12}-x_{35} x_{45} x_{12}+2
   x_{45} x_{12}-x_{12}
   \Big) \,,
\end{align}
and $\mK_5$ has been shown previously. Plugging these into Eq.~\eqref{eq:W_n1_odd}, we readily obtain the letter $W_{5,4}$.

The letter $W_{N,N-1}$ with even $N$ is given by
\begin{equation}
W_{N,N-1} = \frac{\widetilde{B}_{N}-\sqrt{-\widetilde{G}_{N} \mK_{N-1}}}{\widetilde{B}_{N}+\sqrt{-\widetilde{G}_{N} \mK_{N-1}}} \, .
\label{eq:W_n1_even}
\end{equation}
We give an example in topology D. Consider the contribution from the 3-point integral $F_{1,0,1,1,0}$ to the derivative of the 4-point integral $F_{1,1,1,1,0}$. The Gram determinants we need are:
\begin{align}
\widetilde{G}_4 &= G(l,p_1-p_3, -p_4, p_3+p_4)\big|_{\bm{z} = \bm{0}} =\frac{1}{16} (x_{13}-1)^2(x_{34}-4)x_{34} \,, \nonumber \\
\widetilde{B}_4 &= G(l,p_1-p_3, -p_4 ;p_3+p_4,p_1-p_3, -p_4)\big|_{\bm{z} = \bm{0}} \nonumber \\
&=\frac{1}{8} (x_{13}-1)(x_{34}-x_{34} x_{25}+2 x_{25}+x_{13} x_{34}) \,, \nonumber \\
\mK_3 &= G(p_1-p_3, -p_4) = -\frac{1}{4} \left(x_{13}^2+x_{25}^2-2 x_{25} x_{13} -2 x_{13}-2 x_{25} +1\right) .
\end{align}
We can then easily obtain the letter $W_{4,3}$ by plugging the above into Eq.~(\ref{eq:W_n1_even}).

The letter $W_{N,N-2}$ with odd $N$ is given by
\begin{equation}
W_{N,N-2} = \frac{C_N-\sqrt{-\mK_N \mK_{N-2}}}{C_N+\sqrt{-\mK_N \mK_{N-2}}} \, ,
\label{eq:W53}
\end{equation}
where
\begin{equation}
C_N = G(q_1,\ldots,q_{E-2},q_{E-1};\, q_1,\ldots,q_{E-2},q_{E-1}+q_E) \, .
\end{equation}
We give an example in topology A for $t\bar{t}H$ production. Consider $F_{1,1,0,1,0}$'s contribution to $F_{1,1,1,1,1}$, we need 
\begin{align}
C_5 &= G(-p_3, p_1+p_2, -p_4; -p_3, p_1+p_2, -p_2) \nonumber \\
&=\frac{1}{8} \Big(x_{45} x_{12} x_h-x_{12}^2 x_h-2 x_{13} x_{12} x_h+x_{12} x_h-x_{13} x_{12}^2+x_{24} x_{12}^2+2 x_{13} x_{12} \nonumber \\
&+x_{12}-2 x_{24} x_{12}+x_{13} x_{35} x_{12}+x_{13}
   x_{45} x_{12}-2 x_{24} x_{45} x_{12}+x_{35}
   x_{45} x_{12} \nonumber \\
   &+x_{24} x_{45}^2-x_{35} x_{45}^2-x_{13} x_{45}-2
   x_{24} x_{45}+x_{13} x_{35}
   x_{45}+x_{13}-x_{13} x_{35} \nonumber \\
   &+x_{35} x_{45}+x_{45}+x_{24}-2 x_{45} x_{12}-1 \Big) \,, \nonumber \\
\mK_3 &= G(-p_3,p_1+p_2) = -\frac{1}{4} \left(x_{12}^2+x_{45}^2-2 x_{45} x_{12}-2 x_{12}-2x_{45}+1\right) ,
\end{align}
and $\mK_5$ is already given before.

For the letter $W_{N,N-2}$ with even $N$, there are two possible cases. In the first case, both the two $(N-1)$-point integrals between $g_N$ and $g_{N-2}$ are masters, and the letter is given by
\begin{equation}
W_{N,N-2} = \frac{\widetilde{D}_{N} - \sqrt{-\widetilde{G}_{N} \widetilde{G}_{N-2}}}{\widetilde{D}_{N} + \sqrt{-\widetilde{G}_{N} \widetilde{G}_{N-2}}} \, ,
\label{eq:W42}
\end{equation}
where $\widetilde{D}_{N} = D_N(\bm{0})$ and
\begin{equation}
D_N(\bm{z}) = G(k,q_1,\ldots,q_{E-1};k,q_1,\ldots,q_{E-1}+q_E) \, .
\end{equation}
As an example, we show the letter in the contribution from $F_{1,0,0,0,1}$ to $F_{1,0,1,1,1}$ in topology D. The Gram determinants are
\begin{align}
\widetilde{G}_2 &= G(l,p_1+p_2-p_3-p_4)\big|_{\bm{z}=\bm{0}} = \frac{1}{4}x_h (4-x_h) \,, \nn \\
\widetilde{G}_4 &= G(l,p_1+p_2-p_3-p_4,-p_2+p_4,-p_2)\big|_{\bm{z}=\bm{0}} \nn \\
&=\frac{1}{16}(1-x_{24}) \left(4 x_{13} x_h-4 x_h-x_{24} x_{25}^2+x_{25}^2-4
   x_{13} x_{25}+4 x_{24} x_{25}\right) , \nn \\
\widetilde{D}_4 &= G(l, p_1 + p_2 - p_3 - p_4, -p_2 + p_4; l, p_1 + p_2 - p_3 - p_4, -p_2)\big|_{\bm{z}=\bm{0}} \nn \\
&=\frac{1}{8}\left(x_{25} x_h-x_{24} x_{25} x_h+2 x_{13} x_h+2
   x_{24} x_h-4 x_h-2 x_{13} x_{25}+2 x_{24}
   x_{25}\right) .
\end{align}
The second possibility is that one (or both) of the two $(N-1)$-point integrals is not a master and can be reduced to lower-point integrals. Here we only show the letter for the case where the integral $g_{N-1}^{(1)}$ with the denominators $z_1,\ldots,z_{N-1}$ is reducible, while the other $(N-1)$-point integral $g_{N-1}^{(2)}$ is a master. In this case, the letter is given by
\begin{equation}
W_{N,N-2} = \frac{\widetilde{G}_{N-2} \, \mK_N}{\mK_{N-1}^{(1)} \, \widetilde{G}_{N-1}^{(2)}} \, ,
\end{equation}
where we use the superscript $(i)$ to label the Gram determinants associated to the integral $g_{N-1}^{(i)}$. As an example, we consider the contribution from $F_{0,1,0,1,0}$ to $F_{0,1,1,1,1}$ in topology D. The Gram determinants are given by
\begin{align}
\widetilde{G}_2 &= G(l+p_1,-p_3-p_4)\big|_{\bm{z}=\bm{0}} = \frac{1}{4}x_{34} (4-x_{34}) \,, \nn \\
\mK_3^{(1)} &= G(-p_3-p_4, p_4) = \frac{1}{4}x_{34} (4-x_{34}) \, , \nn \\
\mK_4 &=G(-p_3-p_4, p_4, p_2-p_4)  \nn \\
&= \frac{1}{4} \left(x_{34}x_{15}-x_{15}^2+x_{24} x_{34} x_{15}-x_{24} x_{34}^2-x_{24}^2 x_{34}+2 x_{24} x_{34}-x_{34}\right)\,, \nn \\
\widetilde{G}_3^{(2)} &= G(l+p_1,-p_3-p_4, p_2)\big|_{\bm{z}=\bm{0}} = -\frac{1}{4}\left(x_{15} - x_{34}\right)^2 \,.
\end{align}
It happens that $\widetilde{G}_2 = \mK_3^{(1)}$ in this case, and the letter is simply given by $W_{4,2}=\mK_4/\widetilde{G}_3^{(2)}$.

We finally consider the letter $W_{5,2}$. Such dependence can only be present if at least one of the 3-point integrals $g_3^{(i)}$ is reducible. In this case $W_{5,2}$ is the same as (or a combination of) $W_{5,3}$ in Eq.~\eqref{eq:W53}.

The full alphabets for all four topologies in $t\bar{t}H$ production are collected in electronic files attached with this paper. With the alphabets at hand, we reconstruct the $d\bm{A}$ matrices which are also attached. These completely fix the differential equations for the master integrals.

\subsection{Boundary conditions and solution to the differential equations}

Given the canonical bases and their differential equations, we still need to know the value of the integrals at some boundary point $\vec{x}_0$. The boundary points should be chosen for each family such that the integrals become simpler. Our strategy is to utilize the spurious singularities in the differential equations. At these points, many UT master integrals vanish; or equivalently, there are further relations among the master Feynman integrals. It turns out that the most difficult boundary conditions are the 5-point integral in topology A and B, and we discuss their determination in the following.

For topology A we choose the boundary point $\vec{x}_0$ to be 
\begin{equation}
x_{12}=2 \,,\quad  x_{13}=0 \,,\quad x_{24}=0 \,,\quad x_{35}=0 \,,\quad x_{45}=0 \,,\quad x_{h}=0 \, .
\end{equation}
At this point, there are only 7 master integrals which need to be determined. They can be chosen as
\begin{equation}
F_{0,0,0,0,1},\, F_{0,1,0,1,0},\, F_{0,1,0,1,1},\, F_{1,0,0,1,1},\, F_{1,1,0,0,1},\, F_{1,1,0,1,0} ,\, F_{1,1,1,1,1} \, .
\end{equation}
Note that all 4-point integrals can be reduced at the boundary. The lower-point integrals can be easily evaluated as a function of $x_{12}$ using Feynman parameters. The 5-point integral is more difficult, and we obtain its boundary value using a small trick.

First of all, the 5-point integral only appears in one UT master integral ($f_{18}$ in Eq.~\eqref{eq:f_A}). This corresponds to $g_5$ in Eq.~\eqref{eq:UTMIs}, which is UV/IR finite. Hence we have
\begin{equation}
b_{18}(\epsilon) \equiv f_{18}(\vec{x}_0,\epsilon) = c_3 \, \epsilon^3 + \mathcal{O}(\epsilon^4) \, .
\end{equation}
To determine the coefficient $c_3$, we exploit the differential equation with respect to $x_{24}$, while keeping the other variables (collectively denoted as $\vec{x}'$) at the boundary:
\begin{equation}
\frac{\partial}{\partial x_{24}} f_{18}(\vec{x}'_0,x_{24},\epsilon) = \epsilon \, \sum_i A_{i}(x_{24}) \, f_i(\vec{x}'_0,x_{24},\epsilon) \, ,
\end{equation}
where the list $A(x_{24})$ is given by
\begin{align}\nonumber
\bigg\{&0,\frac{1}{8 (x_{24}+1)}-\frac{1}{4 x_{24}}+\frac{1}{8 (x_{24}-1)},0,\frac{1}{8 x_{24}}-\frac{1}{8 (x_{24}+1)},0,\frac{1}{8 (x_{24}-1)}-\frac{1}{8 x_{24}},\\ \nonumber
&\frac{1}{8 (x_{24}+1)}-\frac{1}{8 (x_{24}-1)},\frac{1}{4 x_{24}}-\frac{1}{4 (x_{24}+1)},\frac{1}{2 (x_{24}+1)}-\frac{1}{4 x_{24}}-\frac{1}{4 (x_{24}-1)},\\ \nonumber
&\frac{1}{4 (x_{24}-1)}-\frac{1}{4 (x_{24}+1)},\frac{1}{4 (x_{24}+1)}-\frac{1}{4 x_{24}},-\frac{1}{2 (x_{24}+1)},\frac{1}{2 (x_{24}-1/2)}-\frac{1}{2 x_{24}},\\  \nonumber
&-\frac{1}{2 (x_{24}+1)}+\frac{1}{2 \left(x_{24}-1/2\right)}+\frac{1}{2 x_{24}}-\frac{1}{2 (x_{24}-1)},\frac{1}{2 \left(x_{24}-1/2\right)}-\frac{1}{2 (x_{24}+1)},\\
&\frac{1}{2 x_{24}}-\frac{1}{2 \left(x_{24}-1/2\right)},\frac{1}{2 (x_{24}+1)}-\frac{1}{2 (x_{24}-1/2)},\frac{1}{x_{24}+1}-\frac{1}{x_{24}-1/2} \bigg\} \, .
\end{align}

The differential equation is singular at $x_{24}=1/2$. On the other hand, this is not a genuine singularity of the integrals. Hence, we know that the expansion coefficients $f_{18}^{(n)}(\vec{x}'_0,1/2)$ must be finite for all $n$. Solving the differential equation gives
\begin{equation}
f_{18}^{(n+1)}(\vec{x}'_0,1/2) = \sum_j \int_0^{1/2} dt \, A_{j}(t) \, f_j^{(n)}(\vec{x}'_0,t) + f^{(n+1)}_{18}(\vec{x}'_0,0) \, ,
\end{equation}
It is then required that
\begin{equation}
\lim_{t \to 1/2} \sum_j A_{j}(t) \, f_j^{(n)}(\vec{x}'_0,t) = \text{finite} \, ,
\end{equation}
which allows us to determine $f_{18}^{(n)}(\vec{x}'_0,1/2)$, and in turn $f^{(n)}_{18}(\vec{x}'_0,0)$. The results are expressed as GPLs of the argument $1/2$ (with indices $0$, $1/2$ and $\pm 1$), which can be converted to polylogarithms according to \cite{Frellesvig:2016ske}. The result for $c_3$ is given by
\begin{align}
c_3 &= -\frac{1}{2} \, \mathrm{Li}_3\left(-\frac{1}{2}\right) + \frac{i \pi}{4}  \left[ 2 \mathrm{Li}_2\left(-\frac{1}{2}\right) - \log^2(2) - 2\log(2) \log \left( \frac{3}{2} \right) + \frac{\pi^2}{6} \right] \nonumber
\\
&-\frac{1}{2} \mathrm{Li}_2\left(-\frac{1}{2}\right) \log(2) - \frac{29}{48} \zeta(3) + \frac{1}{12} \log^3(2) +\frac{1}{4}\log^2(2) \log \left(\frac{3}{2}\right) + \frac{\pi^2}{12} \log \left(\frac{3}{2}\right) .
\end{align}

We now turn to topology B. There are 20 master integrals in this family, and only $f_{20}$ involves the 5-point integral $F_{1,1,1,1,1}$. We choose the boundary point $\vec{x}_0$ to be
\begin{equation}
x_{13}=0 \,,\quad  x_{14}=0 \,,\quad x_{24}=0 \,,\quad x_{25}=0 \,,\quad x_{35}=0 \,,\quad x_{h}=0 \, .
\end{equation}
The lower-point integrals at the boundary can be easily calculated. We again need to determine the order $\epsilon^3$ coefficient of $b_{20}(\epsilon) \equiv f_{20}(\vec{x}_0,\epsilon)$. We now exploit the differential equation with respect to $x_h$ (while keeping the other variables at the boundary), and use the fact that $F_{1,1,1,1,1}$ is reducible at the point $x_h = -1$. The boundary condition $b_{20}(\epsilon)$ can then be obtained by evolving from $x_h = -1$ to $x_h = 0$ using the differential equation. However, the resulting expression contains GPLs with complicated indices and arguments such as
\begin{align}
\left\{ \frac{1}{2} \left(\sqrt{5} + 3 \right), \, \frac{1}{2} \left( \sqrt{5} - 1 \right), \, \frac{1}{2}\left(\sqrt{5}+1 \pm \sqrt{3} \pm i\right) , \, \cdots \right\} .
\end{align}
On the other hand, the only letter in the alphabet that survives at the boundary is $d\log(2)$, and it is clear that the boundary conditions can only contain powers of $\log(2)$ in additional to transcendental constants such as zeta values. At weight 3, the only possibilities are $\left\{ \pi^2 \log(2), \zeta_3, \log^3(2)\right\}$ (there can be no imaginary part at the boundary we choose). We use \texttt{GiNaC} to compute the GPLs at the boundary to very high precision, and then use the PSLQ algorithm \cite{pslq} implemented in \texttt{PolyLogTools} \cite{Duhr:2019tlz} to fit the rational coefficients. Finally we find in topology B that
\begin{align}
b_{20}(\epsilon) = \frac{\epsilon^3}{8} \left(\pi^2\log(2) - \frac{13}{2}\zeta_3\right) + \mathcal{O}(\epsilon^4) \, .
\end{align} 

With the boundary conditions at hand, we are now ready to solve the differential equations to get the values of the master integrals at any phase-space point. This step can be done analytically for not-so-complicated integrals. In general, the analytic form is not always easy to obtain, especially for integrals involving many square roots in their alphabets. In these cases, we employ the program package \texttt{DiffExp} \cite{Hidding:2020ytt}, which can take a set of $\epsilon$-form differential equations and compute the solutions numerically using series expansions along a path in the phase space. For a balance between computation speed and accuracy, we aim at results with a relative precision of $10^{-10}$. We cross-check the numeric results for the master integrals with analytic expressions whenever the latter are available. In phase-space regions where the sector decomposition method can get reasonably accurate results, we also cross-check our results to those of \texttt{FIESTA} \cite{Smirnov:2021rhf} and \texttt{pySecDec} \cite{Borowka:2017idc, Borowka:2018goh}, and find agreement within the accuracy. Assembling the integrals into the one-loop amplitudes, we obtain the amplitudes up to $\mathcal{O}(\epsilon^1)$. These serve as an important building block in the two-loop IR divergences we are going to show in the next Section. We have compared the amplitudes up to $\mathcal{O}(\epsilon^0)$ against the results of the program packages \texttt{GoSam} \cite{Cullen:2014yla} and \texttt{OpenLoops} \cite{Buccioni:2019sur, Ossola:2007ax, vanHameren:2010cp, Denner:2016kdg}, and find complete agreements.\footnote{Note that we have been working in conventional dimensional regularization (CDR), where everything lives in $d$ dimensions; while \texttt{GoSam} and \texttt{OpenLoops} implement the 't Hooft-Veltman (HV) scheme, where external legs stay in $4$ dimensions. Our comparisons have taken into account this difference.}

\section{Numeric results and summary}
\label{sec:res}

We now come to the main results of this paper, namely the predictions for the IR poles in the two-loop amplitudes for $t\bar{t}H$ production. In practice, it is more convenient to show the interference between the two-loop amplitudes with the tree-level ones, which is of phenomenological interest. We decompose the color- and spin-summed interference terms into several color coefficients according to \cite{Czakon:2008zk}:
\begin{align}
\label{eq:color_dec}
   2\,{\rm Re} \Braket{{\cal M}_{q}^{(0)}|{\cal M}_{q}^{(2)}}
   &= 2(N^2-1)\,\bigg( N^2 A^q + B^q + \frac{1}{N^2}\,C^q 
    + N n_l\,D_l^q + N n_h\,D_h^q \nonumber
    \\
   &+ \frac{n_l}{N}\,E_l^q 
    + \frac{n_h}{N}\,E_h^q + n_l^2 F_l^q + n_l n_h\,F_{lh}^q 
    + n_h^2 F_h^q \bigg) \,, \nonumber
    \\
   2\,{\rm Re} \Braket{{\cal M}_{g}^{(0)}|{\cal M}_{g}^{(2)}}
   &= (N^2-1)\,\bigg( N^3 A^g + N\,B^g + \frac{1}{N}\,C^g 
    + \frac{1}{N^3}\,D^g \nonumber \\
   &+ N^2 n_l\,E_l^g + N^2 n_h\,E_h^g 
    + n_l\,F_l^g + n_h\,F_h^g + \frac{n_l}{N^2}\,G_l^g 
    + \frac{n_h}{N^2}\,G_h^g \nonumber\\
   &+ N n_l^2 H_l^g + N n_l n_h\,H_{lh}^g 
    + N n_h^2 H_h^g + \frac{n_l^2}{N}\,I_l^g 
    + \frac{n_l n_h}{N}\,I_{lh}^g + \frac{n_h^2}{N}\,I_h^g \bigg) \,.
\end{align}

In Tables~\ref{tab:num1}, \ref{tab:num2}, \ref{tab:num3} and \ref{tab:num4} we list the numeric values of the IR poles as color coefficients at four representative phase-space points. The first point corresponds to the bulk region $M_{t\bar{t}H} \sim 550$~GeV where the differential cross sections are large. The second and third points are in the high energy region where $M_{t\bar{t}H} \sim 2$~TeV. At the second point, the Higgs boson and the top/anti-top quarks are all moderated boosted. At the third point, the top/anti-top quarks are highly boosted while the Higgs boson is produced at relatively low transverse momentum. Finally, the fourth point is near the production threshold where all final state particles have small energies. These results provide a strong check for a future calculation of the two-loop amplitudes. Even before a full two-loop calculation is available, it is possible to study the amplitudes in various kinematic limits, such as the boosted limit or the threshold limit. Our results in these kinematic regions are therefore  useful to validate such calculations.

Comparing to the case of $t\bar{t}$ production, we observe that there are slight differences in $t\bar{t}H$ production due to the $2 \to 3$ kinematics. In particular, the $1/\epsilon^2$ coefficient of $F_h^g$ and the $1/\epsilon$ coefficients of $G_h^g$ and $I_{lh}^g$ vanish for $\mu = m_t$ in $t\bar{t}$ production. However, they are all non-zero in the case of $t\bar{t}H$ production.

\vspace{1ex}

In summary, we have calculated the two-loop infrared divergences in $t\bar{t}H$ production at hadron colliders. To do that we have employed the universal anomalous dimensions obtained in \cite{Ferroglia:2009ep, Ferroglia:2009ii}. We compute the one-loop amplitudes in dimensional regularization up to order $\epsilon^1$, which are important building blocks in the two-loop IR structure. We show the numeric results for the two-loop IR poles at several representative phase-space points. These serve as references for future calculations at this order.

The result in this work is an important part of the ongoing efforts towards the $t\bar{t}H$ cross sections at NNLO. The one-loop amplitudes calculated in this work can be easily extended to order $\epsilon^2$, which are essential ingredients in the NNLO cross sections. It is interesting to study in more detail the behavior of the IR divergences  in the high-energy boosted limit and the low-energy threshold limit, where the amplitudes admit further factorization properties. We leave these to future investigations.

\vspace{3ex}

\textbf{Acknowledgment.} This work was supported in part by the National Natural Science Foundation of China under Grant No. 11975030, 11635001 and 11925506. The research of G. Wang was supported in part by the International Postdoctoral Exchange Fellowship Program (No. PC2021066) from China Postdoctoral Council.

\begin{table}
\begin{center}
\begin{tabular}{||l|r|r|r|r||}
\hline
\hline
&\multicolumn{1}{c|}{$\epsilon^{-4}$}&\multicolumn{1}{c|}{$\epsilon^{-3}$}&\multicolumn{1}{c|}{$\epsilon^{-2}$}&\multicolumn{1}{c|}{$\epsilon^{-1}$}\\
\hline
\hline
$A^{g}$&$17.37022326$&$6.277797530$&$-162.1830217$&$559.8062598$\\
$B^{g}$&$-32.49510001$&$-34.75486260$&$-624.1343773$&$3901.332369$\\
$C^{g}$&&$-9.463444735$&$-54.41556200$&$-497.5350517$\\
$D^{g}$&&&$143.6321997$&$-578.4857199$\\
$E^{g}_l$&&$-20.26526047$&$46.54471184$&$-10.69967085$\\
$E^{g}_h$&&&$-24.23013938$&$79.68650479$\\
$F^{g}_l$&&$37.91095001$&$-74.94866603$&$71.66904977$\\
$F^{g}_h$&&&$43.70151160$&$-132.3384924$\\
$G^{g}_l$&&&$4.731722368$&$85.25318119$\\
$G^{g}_h$&&&&$6.363526190$\\
$H^{g}_l$&&&$3.860049613$&$-10.52987601$\\
$H^{g}_{lh}$&&&&$8.076713126$\\
$H^{g}_h$&&&&\\
$I^{g}_l$&&&$-7.221133335$&$19.49234494$\\
$I^{g}_{lh}$&&&&$-14.56717053$\\
$I^{g}_h$&&&&\\
\hline
\hline
$A^{q}$&$2.390051823$&$15.03938540$&$0.597121534$&$-34.95784899$\\
$B^{q}$&$-4.780103646$&$-22.69017086$&$49.54607207$&$106.0851578$\\
$C^{q}$&$2.390051823$&$7.650785464$&$-186.5751188$&$-21.39439443$\\
$D^{q}_l$&&$-2.390051823$&$0.308675876$&$-6.605875838$\\
$D^{q}_h$&&&$6.244349191$&$4.860387981$\\
$E^{q}_l$&&$2.390051823$&$1.610219156$&$77.52356965$\\
$E^{q}_h$&&&$-6.244349191$&$19.76269918$\\
$F^{q}_l$&&&&\\
$F^{q}_{lh}$&&&&\\
$F^{q}_h$&&&&\\
\hline
\hline
\end{tabular}
\caption{\label{tab:num1}IR poles decomposed as color coefficients for the phase-space point $x_{12}=10$, $x_{13}=-1339/920$, $x_{14}=-2269/465$, $x_{23}=-1951/620$, $x_{24}=-1803/1810$ and $x_{34}=5$.}
\end{center}
\end{table}

\begin{table}
\begin{center}
\begin{tabular}{||l|r|r|r|r||}
\hline
\hline
&\multicolumn{1}{c|}{$\epsilon^{-4}$}&\multicolumn{1}{c|}{$\epsilon^{-3}$}&\multicolumn{1}{c|}{$\epsilon^{-2}$}&\multicolumn{1}{c|}{$\epsilon^{-1}$}\\
\hline
\hline
$A^{g}$&$0.442494477$&$-2.286065211$&$7.804154518$&$-14.93664617$\\
$B^{g}$&$-0.329946796$&$2.483933930$&$-12.80098478$&$40.48995763$\\
$C^{g}$&&$-0.616487423$&$3.068126197$&$-4.327403572$\\
$D^{g}$&&&$0.537127892$&$-1.480913410$\\
$E^{g}_l$&&$-0.516243557$&$2.169405979$&$-5.085548112$\\
$E^{g}_h$&&&$0.508849414$&$-1.628737923$\\
$F^{g}_l$&&$0.384937929$&$-2.005547862$&$4.510173332$\\
$F^{g}_h$&&&$0.014174937$&$-0.068068801$\\
$G^{g}_l$&&&$0.308243712$&$-0.583550524$\\
$G^{g}_h$&&&&$0.013242545$\\
$H^{g}_l$&&&$0.098332106$&$-0.188480542$\\
$H^{g}_{lh}$&&&&$-0.169616471$\\
$H^{g}_h$&&&&\\
$I^{g}_l$&&&$-0.073321510$&$0.139962106$\\
$I^{g}_{lh}$&&&&$-0.004724979$\\
$I^{g}_h$&&&&\\
\hline
\hline
$A^{q}$&$0.034956859$&$-0.070860165$&$0.725834121$&$-4.420336439$\\
$B^{q}$&$-0.069913719$&$0.161141447$&$0.296220955$&$-4.287029013$\\
$C^{q}$&$0.034956859$&$-0.090281281$&$-2.514617229$&$12.94684804$\\
$D^{q}_l$&&$-0.034956859$&$-0.007860626$&$0.513170082$\\
$D^{q}_h$&&&$-0.090618320$&$0.439074083$\\
$E^{q}_l$&&$0.034956859$&$0.065474496$&$0.598204254$\\
$E^{q}_h$&&&$0.090618320$&$0.588540101$\\
$F^{q}_l$&&&&\\
$F^{q}_{lh}$&&&&\\
$F^{q}_h$&&&&\\
\hline
\hline
\end{tabular}
\caption{\label{tab:num2}IR poles decomposed as color coefficients for the phase-space point $x_{12}=130$, $x_{13}=-1831/131$, $x_{14}=-73031/850$, $x_{23}=-34901/1200$, $x_{24}=-1219/77$, $x_{34}=58/3$.}
\end{center}
\end{table}

\begin{table}
\begin{center}
\begin{tabular}{||l|r|r|r|r||}
\hline
\hline
&\multicolumn{1}{c|}{$\epsilon^{-4}$}&\multicolumn{1}{c|}{$\epsilon^{-3}$}&\multicolumn{1}{c|}{$\epsilon^{-2}$}&\multicolumn{1}{c|}{$\epsilon^{-1}$}\\
\hline
\hline
$A^{g}$&$4.597428592$&$-20.35262961$&$66.02325575$&$-113.3974897$\\
$B^{g}$&$-5.840407709$&$39.52398289$&$-243.0196664$&$1108.648507$\\
$C^{g}$&&$-19.61195627$&$106.1882352$&$-329.5335695$\\
$D^{g}$&&&$-2.044016704$&$-83.90927532$\\
$E^{g}_l$&&$-5.363666690$&$20.78482539$&$-47.35659232$\\
$E^{g}_h$&&&$-5.148949659$&$31.29588021$\\
$F^{g}_l$&&$6.813808993$&$-33.36630765$&$82.76573188$\\
$F^{g}_h$&&&$6.957350630$&$-49.77229523$\\
$G^{g}_l$&&&$9.805978134$&$-25.38466647$\\
$G^{g}_h$&&&&$11.68131260$\\
$H^{g}_l$&&&$1.021650798$&$-1.939840823$\\
$H^{g}_{lh}$&&&&$1.716316553$\\
$H^{g}_h$&&&&\\
$I^{g}_l$&&&$-1.297868380$&$2.506852724$\\
$I^{g}_{lh}$&&&&$-2.319116877$\\
$I^{g}_h$&&&&\\
\hline
\hline
$A^{q}$&$0.069015626$&$-0.044110740$&$1.002738019$&$-5.708313785$\\
$B^{q}$&$-0.138031252$&$-0.076438688$&$3.264542288$&$-23.22228179$\\
$C^{q}$&$0.069015626$&$0.120549428$&$-6.917629449$&$27.16467952$\\
$D^{q}_l$&&$-0.069015626$&$-0.055377799$&$0.896831788$\\
$D^{q}_h$&&&$-0.255394642$&$1.035559308$\\
$E^{q}_l$&&$0.069015626$&$0.236793170$&$1.543923421$\\
$E^{q}_h$&&&$0.255394642$&$1.771954886$\\
$F^{q}_l$&&&&\\
$F^{q}_{lh}$&&&&\\
$F^{q}_h$&&&&\\
\hline
\hline
\end{tabular}
\caption{\label{tab:num3}IR poles decomposed as color coefficients for the phase-space point $x_{12}=130$, $x_{13}=-3658/375$, $x_{14}=-88661/845$, $x_{23}=-69551/880$, $x_{24}=-1494/127$, $x_{34}=80$.}
\end{center}
\end{table}

\begin{table}
\begin{center}
\begin{tabular}{||l|r|r|r|r||}
\hline
\hline
&\multicolumn{1}{c|}{$\epsilon^{-4}$}&\multicolumn{1}{c|}{$\epsilon^{-3}$}&\multicolumn{1}{c|}{$\epsilon^{-2}$}&\multicolumn{1}{c|}{$\epsilon^{-1}$}\\
\hline
\hline
$A^{g}$&$20.39809174$&$7.205611610$&$-229.1983000$&$735.4735021$\\
$B^{g}$&$-39.51081921$&$-49.64528951$&$-1599.813719$&$4995.016169$\\
$C^{g}$&&$-2.557923131$&$68.18542213$&$-383.4249655$\\
$D^{g}$&&&$561.6811569$&$-281.6064799$\\
$E^{g}_l$&&$-23.79777369$&$58.16791902$&$0.550271136$\\
$E^{g}_h$&&&$-22.67402287$&$82.67057746$\\
$F^{g}_l$&&$46.09595574$&$-93.70397937$&$123.4757576$\\
$F^{g}_h$&&&$43.21610978$&$-140.2611745$\\
$G^{g}_l$&&&$1.278961565$&$261.5978733$\\
$G^{g}_h$&&&&$1.398901479$\\
$H^{g}_l$&&&$4.532909274$&$-13.50757087$\\
$H^{g}_{lh}$&&&&$7.558007623$\\
$H^{g}_h$&&&&\\
$I^{g}_l$&&&$-8.780182046$&$25.78984026$\\
$I^{g}_{lh}$&&&&$-14.40536993$\\
$I^{g}_h$&&&&\\
\hline
\hline
$A^{q}$&$3.271301102$&$19.44581603$&$-1.765615783$&$-37.05466798$\\
$B^{q}$&$-6.542602204$&$-21.11590873$&$135.0169810$&$280.8915062$\\
$C^{q}$&$3.271301102$&$1.670092695$&$-506.9630348$&$-592.5984017$\\
$D^{q}_l$&&$-3.271301102$&$1.672441341$&$-10.39048008$\\
$D^{q}_h$&&&$11.20597338$&$20.15057383$\\
$E^{q}_l$&&$3.271301102$&$-1.600297099$&$156.6413165$\\
$E^{q}_h$&&&$-11.20597338$&$-7.387295263$\\
$F^{q}_l$&&&&\\
$F^{q}_{lh}$&&&&\\
$F^{q}_h$&&&&\\
\hline
\hline
\end{tabular}
\caption{\label{tab:num4}IR poles decomposed as color coefficients for the phase-space point $x_{12}=3249/400$, $x_{13}=-1301/790$, $x_{14}=-601/225$, $x_{23}=-113/50$, $x_{24}=-544/445$, $x_{34}=21/5$.}
\end{center}
\end{table}

\clearpage

\appendix

\section{The canonical bases}
\label{app:basis}

\subsection{Topology A}

The canonical basis for topology A is given by
\begin{align}
f_1 &= (1-\epsilon) F_{0,0,0,0,1} \,, \nonumber \\
f_2 &= \frac{(\epsilon -1)\left(x_{24}+1\right) F_{0,0,0,0,1} +2 (1-2 \epsilon ) x_{24}F_{0,0,1,0,1}}{2 \left(x_{24}-1\right)} \,, \nonumber \\
f_3 &= \frac{(\epsilon -1) \left(x_{35}+1\right)F_{0,0,0,0,1} +2 (1-2 \epsilon ) x_{35}F_{0,1,0,0,1} }{2 \left(x_{35}-1\right)} \,, \nonumber \\
f_4 &= (2 \epsilon -1) F_{0,1,0,1,0} \,, \nonumber \\
f_5 &= \frac{x_h \left((1- \epsilon ) F_{0,0,0,0,1}-(1-2 \epsilon )
   F_{1,0,0,0,1}\right)}{\sqrt{4-x_h} \sqrt{-x_h}} \,, \nonumber \\
f_6 &= \frac{(\epsilon -1) \left(x_{45}+1\right)F_{0,0,0,0,1} +2 (1-2 \epsilon )
   F_{1,0,0,1,0} x_{45}}{2 \left(x_{45}-1\right)} \,, \nonumber \\
f_7 &= \frac{(\epsilon -1) \left(x_{13}+1\right) F_{0,0,0,0,1} +2 (1-2 \epsilon )
   F_{1,0,1,0,0} x_{13}}{2 \left(x_{13}-1\right)} \,, \nonumber \\
f_8 &= \frac{1}{2} \epsilon \sqrt{-2 x_{12}
   \left(x_{35}+1\right)+x_{12}^2+\left(x_{35}-1\right)^2} F_{0,1,0,1,1} \,, \nonumber \\
f_9 &= \frac{1}{2} \epsilon   \sqrt{-2 x_{45}
   \left(x_h+1\right)+x_{45}^2+\left(x_h-1\right)^2} F_{1,0,0,1,1} \,, \nonumber \\
f_{10} &= \frac{1}{2} \epsilon  \sqrt{-2 x_{13}
   \left(x_{24}+x_h\right)+x_{13}^2+\left(x_{24}-x_h\right)^2} F_{1,0,1,0,1} \,, \nonumber \\
f_{11} &= \frac{1}{2} \epsilon   \sqrt{-2 x_{35}
   \left(x_h+1\right)+x_{35}^2+\left(x_h-1\right)^2}F_{1,1,0,0,1} \,, \nonumber \\
f_{12} &= \frac{1}{2} \epsilon  \sqrt{-2 x_{12}
   \left(x_{45}+1\right)+x_{12}^2+\left(x_{45}-1\right)^2} F_{1,1,0,1,0} \,, \nonumber \\
f_{13} &= -\frac{1}{4} \epsilon   x_{12} \left(x_{24}-1\right)F_{0,1,1,1,1} \,, \nonumber \\
f_{14} &= \frac{1}{4} \epsilon   \left(x_{24}-1\right) \left(x_{45}-1\right)F_{1,0,1,1,1} \,, \nonumber \\
f_{15} &= \frac{1}{4} \epsilon  \sqrt{\left(s_{12} \left(x_h-2\right)-s_{45} s_{35}+s_{35}+s_{45}-1\right)^2-4
   s_{12}^2} F_{1,1,0,1,1} \,, \nonumber \\
f_{16} &= \frac{1}{4} \epsilon  \left(x_{13}-1\right) \left(x_{35}-1\right)F_{1,1,1,0,1} \,, \nonumber \\
f_{17} &= -\frac{1}{4} \epsilon  x_{12} \left(x_{13}-1\right)F_{1,1,1,1,0} \,, \nonumber \\
f_{18} &= c \, F_{1,1,1,1,1} + \cdots \, ,
\label{eq:f_A}
\end{align}
where the expression for $f_{18}$ is too long to be shown here, and we just note that it is the only one that depends on $F_{1,1,1,1,1}$. The full set of expressions are collected in an electronic file attached to this manuscript.

We evaluate the values $\vec{b} \equiv \vec{f}(\vec{x}_0)$ at the boundary point
\begin{align}
\vec{x}_0 = \{x_{12}=2,\  x_{13}=0,\ x_{24}=0,\ x_{35}=0,\ x_{45}=0,\ x_h=0\} \,.
\end{align}
The boundary conditions are:
\begin{align}
b_1&=1 \,,\quad b_2= b_3=b_6=b_7=\frac{1}{2}\,,
b_5=b_{10}=0\,, \nonumber
\\ 
b_4&= -1 +\epsilon  (\log (2) -i \pi ) +\frac{1}{6} \epsilon ^2 \left(4 \pi ^2-3 \log^2 (2) +6 i \pi  \log (2) \right) \nn
\\
&+\frac{1}{6} \epsilon ^3 \left(12 \zeta_3 +2 i \pi ^3+\log^3 (2)-3 i \pi  \log^2 (2) -4 \pi ^2 \log (2) \right) ,\nn
\\
b_8&=b_{12}=\frac{1}{4} \epsilon ^2 (\log (2) -2 i \pi ) \log (2) \nn
\\
&+ \frac{1}{24} \epsilon ^3 \left(3 \zeta_3 -2 i \pi ^3-6 \log^3 (2)+18 i \pi  \log^2 (2) +8 \pi ^2 \log (2) \right) , \nn
\\ 
b_9&=b_{11}=-\frac{\pi ^2 \epsilon ^2}{12}-\frac{3}{2}  \zeta_3 \epsilon ^3 \,, \nn
\\ 
b_{13}&=b_{17}=-\frac{3}{8} +\frac{1}{4} \epsilon  (\log (2) -i \pi ) +\frac{\pi ^2 \epsilon ^2}{6} + \frac{1}{48} \epsilon ^3 \left(27 \zeta_3 +2 i \pi ^3-4 \log^3 (2) +12 i \pi  \log^2 (2) \right) , \nn
\\ 
b_{14}&=b_{16}=\frac{1}{8} -\frac{\pi ^2 \epsilon ^2}{24} -\frac{3}{4} \zeta_3 \epsilon ^3 \,, \nn
\\ 
b_{15} &= \frac{1}{12} \epsilon ^2 \left[ -\pi ^2-3 \log^2(2) +6 i \pi  \log( 2) \right] \nn
\\
&+ \frac{1}{24} \epsilon ^3 \left[-39 \zeta_3+2 i \pi ^3+6 \log^3(2) -18 i \pi  \log^2(2) -8 \pi ^2 \log(2) \right] , \nn
\\ 
b_{18} &= \epsilon ^3 \Bigg\{ -\frac{1}{2} \, \mathrm{Li}_3\left(-\frac{1}{2}\right) + \frac{i \pi}{4}  \left[ 2 \mathrm{Li}_2\left(-\frac{1}{2}\right) - \log^2(2) - 2\log(2) \log \left( \frac{3}{2} \right) + \frac{\pi^2}{6} \right] \nonumber
\\
&-\frac{1}{2} \mathrm{Li}_2\left(-\frac{1}{2}\right) \log(2) - \frac{29}{48} \zeta(3) + \frac{1}{12} \log^3(2) +\frac{1}{4}\log^2(2) \log \left(\frac{3}{2}\right) + \frac{\pi^2}{12} \log \left(\frac{3}{2}\right) \bigg\} \, .
\end{align}

\subsection{Topology B}

The canonical basis for topology B is given by
\begin{align}
f_{1} &= (1-\epsilon) F_{0,0,0,0,1} \,, \nn \\
f_2 &= \frac{(\epsilon -1) \left(x_{24}+1\right)F_{0,0,0,0,1} +2 (1-2 \epsilon )
   x_{24} F_{0,0,1,0,1}}{2 \left(x_{24}-1\right)} \,, \nn \\
f_3 &= \frac{(\epsilon -1)\left(x_{35}+1\right) F_{0,0,0,0,1} +2 (1-2 \epsilon ) x_{35}
   F_{0,1,0,0,1}}{2 \left(x_{35}-1\right)}\,, \nn \\
f_4 &= \frac{(\epsilon -1) \left(x_{14}+1\right) F_{0,0,0,0,1} +2 (1-2 \epsilon )x_{14}
   F_{0,1,0,1,0} }{2 \left(x_{14}-1\right)} \,, \nn \\
f_5 &= \frac{x_h \left((1- \epsilon ) F_{0,0,0,0,1}- (1-2 \epsilon )
   F_{1,0,0,0,1}\right)}{\sqrt{4-x_h} \sqrt{-x_h}} \,, \nn \\
f_6 &= \frac{x_{25} \left((1- \epsilon ) F_{0,0,0,0,1}- (1-2 \epsilon )
   F_{1,0,0,1,0}\right)}{\sqrt{4-x_{25}} \sqrt{-x_{25}}} \,, \nn \\
f_7 &= \frac{(\epsilon -1)\left(x_{13}+1\right) F_{0,0,0,0,1} +2 (1-2 \epsilon ) x_{13}
   F_{1,0,1,0,0}}{2 \left(x_{13}-1\right)} \,, \nn \\
f_8 &= -\frac{1}{2} \epsilon \left(x_{24}-1\right) F_{0,0,1,1,1} \,, \nn \\
f_9 &= \frac{1}{2} \epsilon \left(x_{35}-x_{14}\right) F_{0,1,0,1,1} \,, \nn \\
f_{10} &= \frac{1}{2} \epsilon \left(x_h-x_{25}\right) F_{1,0,0,1,1} \,, \nn \\
f_{11} &= \frac{1}{2} \epsilon \sqrt{-2 x_{13}
   \left(x_{24}+x_h\right)+x_{13}^2+\left(x_{24}-x_h\right){}^2} F_{1,0,1,0,1} \,, \nn \\
f_{12} &= \frac{1}{2} \epsilon \sqrt{-2 x_{13}
   \left(x_{25}+1\right)+x_{13}^2+\left(x_{25}-1\right){}^2} F_{1,0,1,1,0} \,, \nn \\
f_{13} &= \frac{1}{2} \epsilon \sqrt{-2 x_{35}
   \left(x_h+1\right)+x_{35}^2+\left(x_h-1\right){}^2} F_{1,1,0,0,1} \,, \nn \\
f_{14} &= \frac{1}{2} \epsilon \sqrt{-2 x_{14}
   \left(x_{25}+1\right)+x_{14}^2+\left(x_{25}-1\right){}^2} F_{1,1,0,1,0} \,, \nn \\
f_{15} &= \frac{1}{4} \epsilon \left(x_{14}-1\right) \left(x_{24}-1\right) F_{0,1,1,1,1} \,, \nn \\
f_{16} &= \frac{1}{4} \epsilon \sqrt{1-x_{24}} \sqrt{-4 x_{13}
   \left(x_{25}-x_h\right)-x_{24} \left(x_{25}-4\right) x_{25}+x_{25}^2-4 x_h} F_{1,0,1,1,1} \,, \nn \\
f_{17} &= \frac{1}{4} \epsilon \sqrt{x_h-x_{14} x_h+x_{25} x_{35}-x_{25}}
   \sqrt{4 x_{14} -x_{14} x_h +x_{25} x_{35}-x_{25}-4 x_{35}+x_h} F_{1,1,0,1,1} \,, \nn \\
f_{18} &= \frac{1}{4} \epsilon \left(x_{13}-1\right) \left(x_{35}-1\right) F_{1,1,1,0,1} \,, \nn \\
f_{19} &= \frac{1}{4} \epsilon \left(x_{13}-1\right) \left(x_{14}-1\right) F_{1,1,1,1,0} \,, \nn \\
f_{20} &= c \, F_{1,1,1,1,1} + \cdots \, .
\end{align}
The boundary point is chosen as
\begin{align}
\vec{x}_0 = \{x_{13}=0,\, x_{14}=0,\, x_{24}=0, \, x_{25}=0, \, x_{35}=0,\, x_h=0\} \,.
\end{align}
The boundary condition $\vec{b} = \vec{f}(\vec{x}_0,\epsilon)$ is
\begin{align}
b_1&=1 \,, \quad 
b_2=
b_3=
b_4=b_7=\frac{1}{2} \,, \quad 
b_5=
b_6=b_9=b_{10}=b_{11}=b_{16}=b_{17}=0 \,, \nn
\\
b_8&=b_{12}=b_{13}=b_{14}= -\frac{\pi ^2 \epsilon ^2}{12} -\frac{3}{2}  \zeta_3 \epsilon ^3 \,, \nn
\\ 
b_{15}&=b_{18}= \frac{1}{8} -\frac{\pi ^2 \epsilon ^2}{24} -\frac{3}{4}  \zeta_3 \epsilon ^3 \,, \nn
\\ 
b_{19}&= \frac{1}{4} -\frac{\pi ^2 \epsilon ^2}{6} + \epsilon ^3 \left(-\frac{23 \zeta_3}{8}-\frac{1}{4} \pi ^2 \log (2) \right) , \nn
\\ 
b_{20}&=\epsilon ^3 \left(\frac{1}{8} \pi ^2 \log (2)-\frac{13 \zeta_3}{16}\right) .
\end{align}

\subsection{Topology C}

The canonical basis for topology C is chosen as
\begin{align}
f_1 &= (1-\epsilon) F_{0,0,0,0,1} \,, \nn \\
f_2 &= \frac{x_{12} \left((1- \epsilon ) F_{0,0,0,0,1}- (1-2 \epsilon )
   F_{0,0,1,0,1}\right)}{\sqrt{4-x_{12}} \sqrt{-x_{12}}} \,, \nn \\
f_3 &= \frac{(\epsilon -1)\left(x_{35}+1\right) F_{0,0,0,0,1} +2 (1-2 \epsilon )x_{35}
   F_{0,1,0,0,1} }{2 \left(x_{35}-1\right)} \,, \nn \\
f_4 &= \frac{(\epsilon -1)\left(x_{24}+1\right) F_{0,0,0,0,1} +2 (1-2 \epsilon )x_{24}
   F_{0,1,0,1,0} }{2 \left(x_{24}-1\right)} \,, \nn \\
f_5 &= \frac{x_h \left((1-\epsilon ) F_{0,0,0,0,1}- (1-2 \epsilon )
   F_{1,0,0,0,1}\right)}{\sqrt{4-x_h} \sqrt{-x_h}} \,, \nn \\
f_6 &= \frac{x_{15} \left((1- \epsilon ) F_{0,0,0,0,1}-(1-2 \epsilon )
   F_{1,0,0,1,0}\right)}{ \sqrt{4-x_{15}} \sqrt{-x_{15}}} \,, \nn \\
f_7 &= \frac{x_{34} \left((1- \epsilon ) F_{0,0,0,0,1}- (1-2 \epsilon )
   F_{1,0,1,0,0}\right)}{ \sqrt{4-x_{34}} \sqrt{-x_{34}}}, \nonumber \\
f_8 &= \frac{1}{2} \epsilon x_{12} F_{0,0,1,1,1} \,, \nn \\
f_9 &= \frac{1}{2} \epsilon \left(x_{35}-x_{24}\right) F_{0,1,0,1,1} \,, \nn \\
f_{10} &= \frac{1}{2} \epsilon \sqrt{-2 x_{12}
   \left(x_{35}+1\right)+x_{12}^2+\left(x_{35}-1\right){}^2} F_{0,1,1,0,1} \,, \nn \\
f_{11} &= -\frac{1}{2} \epsilon  \left(x_{24}-1\right) F_{0,1,1,1,0} \,, \nn \\
f_{12} &= \frac{1}{2} \epsilon \left(x_h-x_{15}\right) F_{1,0,0,1,1} \,, \nn \\
f_{13} &= \frac{1}{2} \epsilon \sqrt{-2 x_{12}
   \left(x_{34}+x_h\right)+x_{12}^2+\left(x_{34}-x_h\right){}^2} F_{1,0,1,0,1} \,, \nn  \\
f_{14} &= \frac{1}{2} \epsilon \left(x_{15}-x_{34}\right) F_{1,0,1,1,0} \,, \nn \\
f_{15} &= \frac{1}{2} \epsilon \sqrt{-2 x_{35}
   \left(x_h+1\right)+x_{35}^2+\left(x_h-1\right){}^2} F_{1,1,0,0,1} \,,  \nn \\
f_{16} &= \frac{1}{2} \epsilon \sqrt{-2 x_{15}
   \left(x_{24}+1\right)+x_{15}^2+\left(x_{24}-1\right){}^2} F_{1,1,0,1,0} \,, \nn \\
f_{17} &= \frac{1}{4} \epsilon \sqrt{-x_{12}} \sqrt{1-x_{24}} \sqrt{x_{12}
   \left(x_{24}-1\right)-4 x_{24}+4 x_{35}} F_{0,1,1,1,1} \,, \nn \\
f_{18} &= \frac{1}{4} \epsilon \sqrt{-x_{12}} \sqrt{4
   \left(x_{15}-x_{34}\right) \left(x_{15}-x_h\right)-x_{12}
   \left(x_{15}-4\right) x_{15}} F_{1,0,1,1,1} \,, \nn \\
f_{19} &= \frac{1}{4} \epsilon \sqrt{-x_{15} \left(x_{35}-1\right)+x_{24}
   \left(x_h-4\right)+4 x_{35}-x_h} \sqrt{\left(x_{24}-1\right) x_h-x_{15}
   \left(x_{35}-1\right)} F_{1,1,0,1,1} \,, \nn \\
f_{20} &= -\frac{1}{4} \epsilon \left(x_{35}-1\right)\sqrt{4-x_{34}} \sqrt{-x_{34}} F_{1,1,1,0,1} 
   \,, \nn \\
f_{21} &= -\frac{1}{4} \epsilon \left(x_{24}-1\right) \sqrt{4-x_{34}}
   \sqrt{-x_{34}} F_{1,1,1,1,0} \,, \nonumber \\
f_{22} &= c \, F_{1,1,1,1,1} + \cdots \, .
\end{align}
The boundary point is chosen as
\begin{align}
\vec{x}_0 = \{x_{12}=0,\, x_{15}=0,\, x_{24}=0, \, x_{34}=0, \, x_{35}=0,\, x_h=0\} \,.
\end{align}
The boundary condition $\vec{b} = \vec{f}(\vec{x}_0,\epsilon)$ is
\begin{align}
b_1&=1 \,,\quad 
b_3=b_4=\frac{1}{2}\,,\nn\\ 
b_2&=
b_5= 
b_6=b_{7}=b_8=b_9=b_{12}=b_{13}=b_{14}=
b_{17}=
b_{18}=
b_{19}=
b_{20}=
b_{21}=
b_{22}=0 \, , \nn \\ 
b_{10}&=b_{11}=b_{15}=b_{16}=-\frac{1}{180} \epsilon ^2 \left(15 \pi ^2 + 270 \zeta_3 \epsilon +7 \pi ^4 \epsilon ^2\right) .
\end{align}

\subsection{Topology D}

The canonical basis for topology D is chosen as
\begin{align}
f_1 &= (1-\epsilon) F_{0,0,0,0,1} \,,\nn\\
f_2 &= \frac{(\epsilon -1)\left(x_{24}+1\right) F_{0,0,0,0,1} +2 (1-2 \epsilon )x_{24}
   F_{0,0,1,0,1} }{2 \left(x_{24}-1\right)} \,,\nn  \\
f_3 &= \frac{x_{15} \left((1- \epsilon ) F_{0,0,0,0,1}- (1-2 \epsilon )
   F_{0,1,0,0,1}\right)}{ \sqrt{4-x_{15}} \sqrt{-x_{15}}} \,,\nn  \\
f_4 &= \frac{x_{34} \left((1- \epsilon ) F_{0,0,0,0,1}- (1-2 \epsilon )
   F_{0,1,0,1,0}\right)}{ \sqrt{4-x_{34}} \sqrt{-x_{34}}} \,,\nn  \\
f_5 &= \frac{x_h \left((1- \epsilon ) F_{0,0,0,0,1}-(1-2 \epsilon )
   F_{1,0,0,0,1}\right)}{ \sqrt{4-x_h} \sqrt{-x_h}} \,,\nn  \\
f_6 &= \frac{x_{25} \left((1- \epsilon ) F_{0,0,0,0,1}- (1-2 \epsilon )
   F_{1,0,0,1,0}\right)}{ \sqrt{4-x_{25}} \sqrt{-x_{25}}} \,,\nn  \\
f_7 &= \frac{(\epsilon -1)\left(x_{13}+1\right) F_{0,0,0,0,1} +2 (1-2 \epsilon ) x_{13}
   F_{1,0,1,0,0}}{2 \left(x_{13}-1\right)} \,,\nn  \\
f_8 &= -\frac{1}{2} \epsilon \left(x_{24}-1\right) F_{0,0,1,1,1} \,,\nn  \\
f_9 &= \frac{1}{2} \epsilon \left(x_{15}-x_{34}\right) F_{0,1,0,1,1} \,,\nn  \\
f_{10} &= \frac{1}{2} \epsilon \sqrt{-2 x_{15}
   \left(x_{24}+1\right)+x_{15}^2+\left(x_{24}-1\right){}^2} F_{0,1,1,0,1} \,,\nn   \\
f_{11} &= \frac{1}{2} \epsilon \left(x_h-x_{25}\right) F_{1,0,0,1,1} \,,\nn  \\
f_{12} &= \frac{1}{2} \epsilon \sqrt{-2 x_{13}
   \left(x_{24}+x_h\right)+x_{13}^2+\left(x_{24}-x_h\right){}^2} F_{1,0,1,0,1} \,,\nn  \\
f_{13} &= \frac{1}{2} \epsilon \sqrt{-2 x_{13}
   \left(x_{25}+1\right)+x_{13}^2+\left(x_{25}-1\right){}^2} F_{1,0,1,1,0} \,,\nn \\
f_{14} &= \frac{1}{2} \epsilon \left(x_h-x_{15}\right) F_{1,1,0,0,1} \,,\nn  \\
f_{15} &= \frac{1}{2} \epsilon \left(x_{25}-x_{34}\right) F_{1,1,0,1,0} \,, \nonumber \\
f_{16} &= -\frac{1}{2} \epsilon  \left(x_{13}-1\right) F_{1,1,1,0,0}\,,\nn  \\
f_{17} &= -\frac{1}{4} \epsilon \left(x_{24}-1\right) \sqrt{4-x_{34}}
  \sqrt{-x_{34}} F_{0,1,1,1,1} \,,\nn \\
f_{18} &= \frac{1}{4} \epsilon \sqrt{1-x_{24}} \sqrt{-4 x_{13}
   \left(x_{25}-x_h\right)-x_{24} \left(x_{25}-4\right) x_{25}+x_{25}^2-4 x_h} F_{1,0,1,1,1} \,,\nn  \\
f_{19} &= \frac{1}{4} \epsilon \sqrt{x_{15} x_{25}-x_{34} x_h} \sqrt{x_{15}
   \left(x_{25}-4\right)-4 x_{25}-x_{34} x_h+4 x_{34}+4 x_h} F_{1,1,0,1,1} \,,\nn  \\
f_{20} &= \frac{1}{4} \epsilon \sqrt{1-x_{13}} \sqrt{-x_{13}
   \left(x_{15}-4\right) x_{15}-4 x_{15} x_{24}+x_{15}^2+4 \left(x_{24}-1\right)
   x_h} F_{1,1,1,0,1} \,,\nn  \\
f_{21} &= -\frac{1}{4} \epsilon \left(x_{13}-1\right) \sqrt{4-x_{34}}
   \sqrt{-x_{34}} F_{1,1,1,1,0} \,, \nn\\
f_{22} &= c \, F_{1,1,1,1,1} + \cdots \, .
\end{align}
The boundary point is chosen as
\begin{align}
\vec{x}_0 = \{x_{13}=0,\, x_{15}=0,\, x_{24}=0, \, x_{25}=0, \, x_{34}=0,\, x_h=0\}\,.
\end{align}
The boundary condition $\vec{b} = \vec{f}(\vec{x}_0,\epsilon)$ is
\begin{align}
b_1&=1\,,\quad  
b_2=b_7=\frac{1}{2}\,,\nn\\
b_3&=
b_4=
b_5= 
b_6=b_{9}=b_{11}=b_{12}=b_{14}=b_{15}=
b_{17}=
b_{18}=
b_{19}=
b_{20}=
b_{21}=
b_{22}=0\,,\nn\\ 
b_8&=b_{10}=b_{13}=b_{16}=-\frac{1}{180} \epsilon ^2 \left(15 \pi ^2 + 270 \zeta_3 \epsilon +7 \pi ^4 \epsilon ^2\right).
\end{align}

\bibliographystyle{JHEP}
\bibliography{references_inspire.bib,references_local.bib}

\end{document}